\documentclass[]{aastex63}

\def\Av{$\rm{A_{V}}$}

\def\Teff{$\rm{T_{eff}}$}

\received{\today}
\revised{\today}
\submitjournal{AJ}

\shorttitle{Angular momentum evolution}
\shortauthors{Pinz\'on et al.}

\begin{document}

\title{Understanding the angular momentum evolution of T Tauri and Herbig Ae/Be stars}

\correspondingauthor{Giovanni Pinz\'on}
\email{gapinzone@unal.edu.co}

\author[0000-0001-9147-3345]{Giovanni Pinz\'on}
\affiliation{Observatorio Astron\'omico Nacional, Facultad de Ciencias, Universidad Nacional de Colombia, Bogot\'a, Colombia}

\author[0000-0001-9797-5661]{Jes\'us Hern\'andez}
\affil{Instituto de Astronom\'{i}a, Universidad Aut\'{o}noma de M\'{e}xico, Ensenada, B.C, M\'{e}xico}

\author[0000-0001-7351-6540]{Javier Serna}
\affil{Instituto de Astronom\'{i}a, Universidad Aut\'{o}noma de M\'{e}xico, Ensenada, B.C, M\'{e}xico}

\author{Alexandra Garc\'ia}
\affiliation{Observatorio Astron\'omico Nacional, Facultad de Ciencias, Universidad Nacional de Colombia, Bogot\'a, Colombia}

\author[0000-0001-6647-862X]{Ezequiel Manzo-Mart\'{\i}nez}
\affiliation{Mesoamerican Centre for Theoretical Physics, Universidad Aut\'onoma de Chiapas, Carretera Zapata Km. 4, Real del Bosque (Ter\'an), Tuxtla Guti\'errez 29040, Chiapas, M\'exico}

\author[0000-0002-1379-4204]{Alexandre Roman-Lopes}
\affiliation{Department of Astronomy - Universidad de La Serena - Av. Juan Cisternas, 1200 North, La Serena, Chile}

\author[0000-0000-0000-0000]{Carlos G. Rom\'an-Z\'u\~niga}
\affil{Instituto de Astronom\'{i}a, Universidad Aut\'{o}noma de M\'{e}xico, Ensenada, B.C, M\'{e}xico}

\author{Maria Gracia Batista}
\affiliation{Observatorio Astron\'omico Nacional, Facultad de Ciencias, Universidad Nacional de Colombia, Bogot\'a, Colombia}

\author{Julio Ram\'irez-V\'elez}
\affil{Instituto de Astronom\'{i}a, Universidad Aut\'{o}noma de M\'{e}xico, Ensenada, B.C, M\'{e}xico}

\author[0000-0001-5832-6933]{Yeisson Osorio}
\affiliation{Instituto de Astrofísica de Canarias, E-38205 La Laguna, Tenerife, Spain}
\affiliation{Departamento de Astrofísica, Universidad de La Laguna (ULL), E-38206 La Laguna, Tenerife, Spain}

\author{Ronald Avenda\~no}
\affiliation{Observatorio Astron\'omico Nacional, Facultad de Ciencias, Universidad Nacional de Colombia, Bogot\'a, Colombia}

\begin{abstract}
We  investigate a sample of 6 Herbig Ae/Be stars  belonging to the Orion OB1 association, as well as  73 low mass objects, members of the $\sigma$ Orionis cluster, in order to explore the angular momentum evolution at  early stages of evolution, and its possible connection  with main-sequence Ap/Bp magnetic stars. Using FIES and HECTOCHELLE spectra, we obtain projected rotational velocities through two independent methods. Individual masses, radii, and ages were computed using evolutionary models, distance, and cluster extinction.  Under the  assumption that similar physical processes operate  in both, T Tauri and Herbig Ae/Be stars, we construct snapshots of the protostar's rotation against mass during the first 10 Myr  with the aid of a rotational model that includes a variable disc  lifetime, changes in the stellar moment of inertia,  a dipolar magnetic field with variable strength, and angular momentum loss through stellar winds powered by accretion.  We use these snapshots, 
as well as the rotational data, to infer  a plausible scenario for the angular momentum evolution.  We find that  magnetic field strengths of a few k$G$ at 3 Myr  are required  to match the rotational velocities  of both groups of stars. Models with masses between 2-3 $M_{\odot}$ display  larger  angular momentum by a factor of $\sim 3$, in comparison to stars of similar spectral types on the main-sequence.  Even though some quantitative estimates on this dramatic decrease with age, for  Ap/Bp magnetic main-sequence stars are presented,  the results obtained 
   for the angular momentum evolution do not explain their low rotation.

%Angular momentum of Ap/Bp stars is on average one tenth of that in HAeBes.  We discuss these results in the context of a scenario in which Ap/Bp stars loss  a quarter of its angular momentum during their PMS stage.

\end{abstract}

Stellar physics Star forming regions
\section{Introduction} \label{sec:intro}

  The angular momentum evolution of pre-main sequence stars (PMS) is mainly affected by internal processes which determine the angular momentum redistribution throughout the stellar interior \citep{macgregor1991,moss1992,garaud2020}, as well as external processes such as stellar winds \citep{hayashi1996,cranmer2008} and the interaction between the star and the protoplanetary disc \citep{armitage1996,romanova2005,matt2005,matt2012,gallet2013,gallet2015,li2014},  all of them natural products of angular momentum conservation during the gravitational collapse of dense cores into protostars \citep{hartmann1998}.  In the current magnetospheric accretion scenario (MA), low mass ( $\le1.5M_{\odot}$)  T Tauri stars (TTs)   have their own magnetic field that truncates the disc at a few stellar radii from the star surface. From such location, gas from the disc with higher angular momentum falls onto the star along the magnetic field lines during the so-called accretion phase. The energy  released  during accretion may excite large fluxes of Alfv\'en waves along open field lines in the magnetosphere, that is, stellar winds, which play an important role in the outward transfer of angular momentum \citep{bohm1994,corcoran1997,matt2008}.  While this scenario describes  angular momentum in TTs, its extrapolation toward intermediate-mass stars  is not straightforward.  The main dilemma is that only a small fraction of intermediate-mass (B and A) stars exhibit significant magnetic fields \citep{wade2007}. However, several spectropolarimetric studies confirm the  presence of weak fields  in a number of Herbig Ae/Be stars \citep{hubrig2006,alecian2008,alecian2013}. In addition, the interplay between stellar accretion, winds and protostellar discs in intermediate-mass stars is still matter of debate. There is evidence suggesting that more massive stars accrete material from their discs in different ways to that predicted by MA theory  \citep{cauley2014,cauley2015}. However, recent  findings indicate that MA may still operate in intermediate-mass pre-main sequence stars such as HQ Tau \citep{pouilly2020}. Furthermore, a trend between the strength of the longitudinal magnetic field and the accretion rate discussed in the work of \cite{hubrig2009b} qualitatively supports the MA scenario for Herbig Ae/Be stars.  \\

  Herbig Ae/Be stars (hereafter HAeBes) are young stellar objects with masses between  (1.5-10 $M_{\odot}$) which share characteristics with TTs  despite differences in their stellar masses and internal structures. Both groups exhibit similar ages, infrared  and ultraviolet excesses above the photosphere, signs of magnetic activity, and some display P-Cygni absorption features originated in strong collimated  winds \citep{hubrig2014,wichittanakom2020}.  Internal structure of HAeBes differs from that of TTs; while the former is mainly radiative, TTs are mostly fully-convective \citep{iben1965}.  This fact has a tremendous impact on the magnetic field generation and situates HAeBes as  excellent candidates for testing the mass dependence of MA scenario, in particular, details about whether they constitute a scaled version of TTs or not. \\
  
  %As both are young stellar objects, HAeBes are  excellent subjects for testing the mass-dependence of MA scenario, in particular details about whether they constitute a scaled version of TTs or not.  }\\

   Despite lacking outer convective zones required to power a solar-type dynamo as observed in TTs \citep{moss1984} almost $\sim$10\% of the HAeBes exhibit predominantely dipolar magnetic fields with strengths  between tens and thousands of $G$ \citep{alecian2013}.  These fields inferred from circular polarization measurements of selected line
profiles are similar to those found in the peculiar (Ap/Bp) main-sequence stars, which are slow rotators
with periods in the range from days to decades  \citep{zorec2012}.  This similarity can be interpreted
as evidence supporting that HAeBes are progenitors of the Ap/Bp stars. It has been suggested that Ap/Bp stars do not undergo significant angular momentum losses during the main-sequence phase and that any loss of angular momentum must occur either in the PMS phase or at the beginning of the main-sequence life, before the stars become observably magnetic \citep{north1998,hubrig2000,kb2006}.   \citet{rosen2012} carried out numerical simulations of the angular momentum evolution of accreting massive stars. They used a magnetically controlled scenario to demonstrate that  magnetic star-disc interaction's torques alone are insufficient to spin down massive young stars. This implies that other sources of angular momentum loss, such as stellar winds, must be taken into account. It also confirms that intermediate-mass stars loss angular momentum during their PMS stage under similar mechanisms than those operating in their lower mass counterparts \citep{auriere2007}.\\

 A  better  understanding of the angular momentum  distribution as a function of the spectral type can be achieved by using the results of multi-epoch spectroscopic large surveys, such as Gaia-ESO\footnote{https://www.cosmos.esa.int/web/gaia} ($\ge10^5$ stars) and SDSS-V\footnote{https://www.sdss.org/future/} ($\ge6\times10^6$ stars) 
. These surveys
 provide stellar parameters, including stellar rotation measurements with high accuracy.  Careful comparisons of projected rotational velocities ($v$sin$i$) from both the literature and synthetic values are highly needed. In this sense, global rotational models of young stars capable to predict trends between rotation, spectral types, accretion, and magnetic field for a large number of objects are necessary.  \\

In this work we use a rotational model to search for  trends between rotation, mass,  stellar magnetic fields and ages of TTs and HAeBes. We implement the magnetospheric accretion model of TTs toward higher and lower stellar masses in order to describe the angular momentum evolution for 
 both mass regimes.  For these purposes, we consider an approximately coeval sample of young stellar objects containing both TTs and HAeBes. We selected  HAeBes members of the young Orion OB1 association, particularly from the subgroups $a$ and $bc$ \citep[1-7 Myr;]{hernandez2005}, and confirmed TTs members from the $\sigma$ Orionis cluster \citep[$\sim$ 3 Myr;][]{hernandez2007}.
We use this sample to study the interplay between rotation, the accreting disc, and stellar winds in intermediate-mass stars.   \\

The paper is organized as follows. In Section 2, we describe the sample selection and  observations. Stellar parameters are presented in Section 3, while in Section 4, we discuss the main features of rotation of TTs and HAeBes. The interplay between winds and accretion in HAeBes is described in Section 5, whereas synthetic $v$sin$i$-mass relations and the angular momentum of HAeBes are discussed in Sections 6 and 7, respectively.  In Section 8, we present a summary and our conclusions.

\section{Sample selection and observations} \label{sec:obs}
 
\subsection{The FIES echelle spectra  }

We selected 5 HAeBe stars located in the Orion OB1 association identified by \citet{hernandez2005} using the following criteria: (1) H$\alpha$
in emission, (2) location in the HAeBe region of the JHK color color diagram, and (3) strong IRAS 12{\micron} fluxes \citep{vieira2003}. Two of the HAeBe stars are located in the Orion OB1a association (age $\sim$7-10 Myr) and the other three are located in the Orion OB1bc association (age $\sim$3-5 Myr). We also included the star HIP26500, which has an uncertain status between classical Be (CBe) star and HAeBe star \citep{hernandez2005}. Since HIP26500 has H$\alpha$ in emission and significant infrared excesses \citep[e.g,][]{chen2016,vioque2018}, we list this object as a HAeBe star. \\

We acquired high-resolution spectra of these six HAeBes (hereafter HAeBe sample) and two comparison stars (HD87737 and HD53244) with the Fibre-fed Echelle Spectrograph (FIES\footnote{Observations made with the Nordic Optical Telescope, operated by the Nordic Optical Telescope Scientific Association at the Observatorio del Roque de los Muchachos, La Palma, Spain, of the Instituto de Astrofisica de Canarias.}) on the Nordic Optical 2.5m telescope (NOT) on January 20th of 2014.  FIES enables a resolution of 68000 in the wavelength range of 3680-7270\AA. Exposure times were 400 seconds. Spectra were processed with the FIES pipeline FIEStools following  typical reduction steps for echelle data such as bias subtraction, flat-field normalization, extraction of the spectra and assignment of wavelengths along the spectra. The stars HIP26752 and HIP25258 are spectroscopic binary systems  \citep[e.g.][]{vioque2018}. We acquired spectroscopic data for the two components of the system HIP25258 which have an angular separation of $\sim$2.3\arcsec. Since the angular separation of the components of the system HIP26752 is smaller than 2\arcsec \citep{wheelwright2010},  we have obtained FIES spectra for the combined stellar system. \\

Table \ref{tab:table1} lists identifier, coordinates, spectral type, effective temperature (\Teff), visual extinction (\Av) and location reported by \citet{hernandez2005},  for our sample. Table \ref{tab:table1} also shows distances estimated from   Gaia-EDR3 parallaxes \citep{gaia2020}, since uncertainties for all our sources are smaller than 5\%, thus we can apply the inverse relation between distance and parallax \citep{bailer2015}. For comparison, we also show the distances estimated by \citet{briceno2019} for the sub-associations  Orion OB1a and Orion OB1b. The star HIP27059 is located in the spatial limit between the Orion OB1a and the Orion OB1b associations and has a distance similar to young stellar populations located in the Orion OB1a association \citep{briceno2019, perez2018}. Thus, it is probable that HIP27059 belongs to the Orion OB1a association instead  to the Orion OB1b, as previously reported by \citet{hernandez2005}.  In Table \ref{tab:table2} we show properties of the template spectra used for radial and rotational measurements which were selected from the radial velocities catalogue of \citet{debruijne2012}.

\subsection{The Hectochelle spectra}

Using spectra obtained with the Hectochelle fiber-fed multiobject echelle spectrograph at the 6.5 m Telescope of the MMT Observatory (MMTO), \citet{hernandez2014} reported radial velocity (RVs), H$\alpha$, and \ion{Li}{1} measurements for 142 stars in the $\sigma$ Orionis cluster. These spectra have a resolution of 34000 with 180 {\AA} of spectral coverage centered at 6625\AA. 
 From this data set, we complemented our study determining projected rotational velocities for  73 confirmed members that exhibit \ion{Li}{1}   in absorption   (hereafter TTs sample; see \S \ref{subsec:rotvel}).

\section{Stellar parameters} \label{sec:stellarparams}

 \subsection{Stellar masses, ages and radii} \label{subsec:masa}

 Based on previous estimations of spectral type and visual extinction given by \cite{hernandez2005} for the HAeBes sample and by \cite{hernandez2014} for TTs sample, we computed effective temperatures ({\Teff}) by interpolating the spectral types  using the standard table given by \cite{pm2013} for pre-main sequence stars.  Subsequently, we determined stellar luminosities based on 2MASS J band, the Gaia EDR3 parallaxes\footnote{the parallaxes uncertainties in Gaia-EDR3 are below 20\% for this sample, thus we can apply this inverse relation \citep{bailer2015}} and the extinction law  of \cite{cardelli1989}.  Finally, we computed stellar masses and ages by comparing the position of the stars in the H-R diagram with evolutionary models through the use of the code \textit{MassAge} (Hern\'andez et al., in preparation). This code generates for each star, 300 artificial points, considering the values and uncertainties of the extinctions, spectral types, J magnitudes and parallaxes. We select the stellar mass and age of the closest theoretical point in the MIST evolutionary model grid for each artificial point.\\
 
\begin{figure}
\plotone{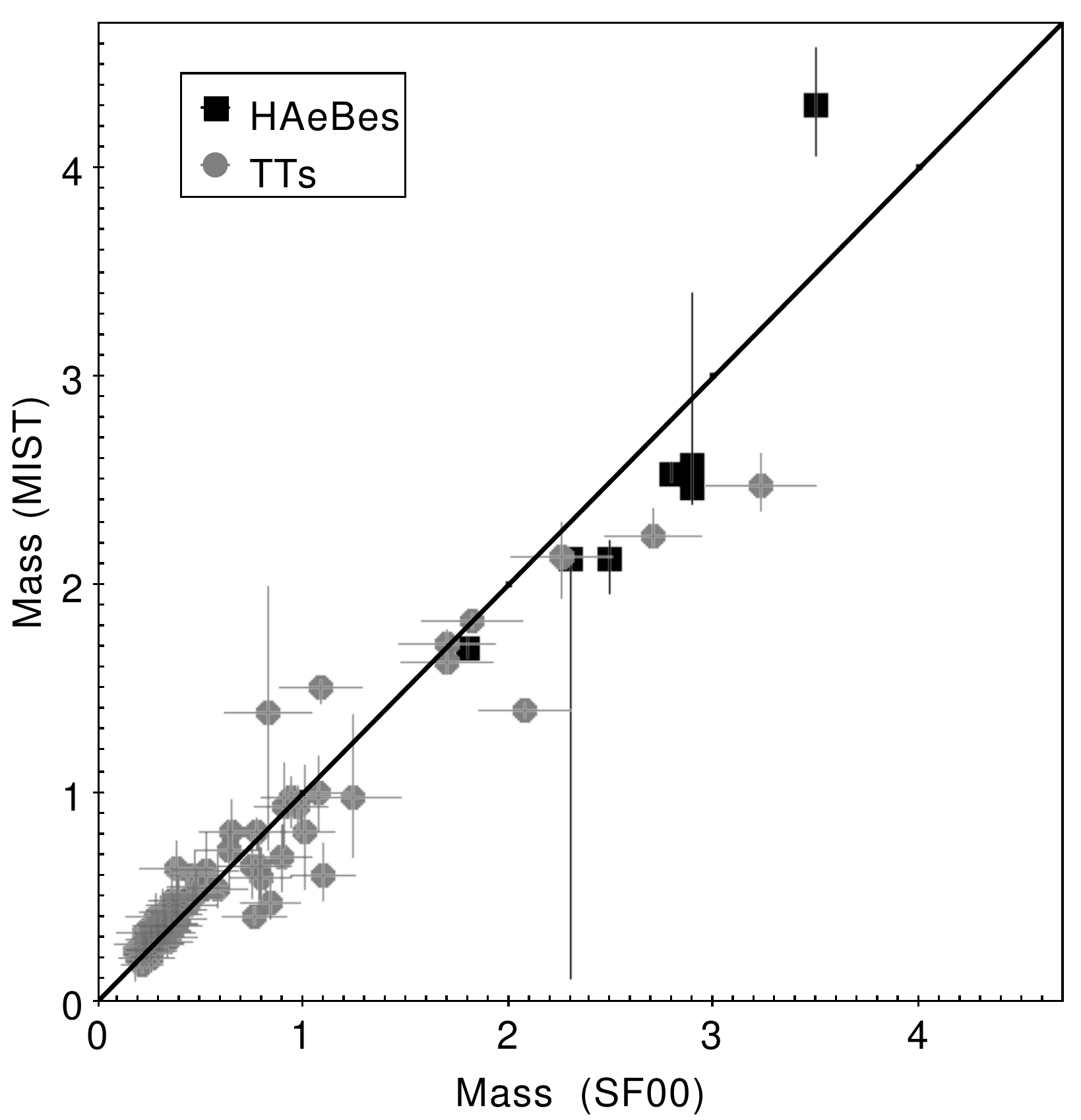}
\caption{  Comparison between the stellar masses derived through \textit{MassAge} code using evolutionary models of MIST \citep{dotter2016} and SF00 models \citep{siess2000} for our sample. TTs are confirmed members of $\sigma-$Orionis with LiI in absorption. The MIST-mass of the HAeBe binary HIP25258 corresponds to the principal component (see in the text). }     
\label{fig:f0}
\end{figure}

 We report the median value of age and mass for each star. The upper and lower limits correspond to the 1-sigma level, where 68\% of the individual results are included. In Figure \ref{fig:f0}, we compare the stellar masses derived from MIST \citep{dotter2016} and  \cite{siess2000}.  In general, stellar masses are the same within the uncertainties. Stellar radii were obtained from the luminosity equation. For the HIP25258 binary system, whose components have similar parallaxes but magnitudes quite different, we estimate stellar parameters for the principal component. Masses, ages, and radii for the TTs are listed in Table 3 and those for the HAeBes in Table \ref{tab:tablehaebes}.

\subsection{Radial velocities} \label{subsec:radvel}

Radial velocities (RVs) for the HAeBes sample were computed using the IRAF task FXCOR, through a cross-correlation function (hereafter, CCF) that compares each star with a synthetic or empirical template with similar spectral type, corrected by radial velocity. Each object's spectrum was cross-correlated with its respective template in the spectral window $4480${\AA}$<\lambda<4600$ {\AA}. Subsequently, the quality of the cross-correlation was measured through the parameter $R$, defined as the S/N of the cross-correlation function \cite{tr1979}. Heliocentric radial velocities in Table \ref{tab:tablehaebes}  are those that give the highest $R$ values.  Uncertainties were computed through the $R$ parameter as $\sigma_v=v_{rad}/(1+R)$.  \\

For five of our HAeBes stars, \cite{alecian2013} measured RVs through fitting of LSD Stokes I profiles obtained with the high-resolution spectropolarimeters ESPaDOnS and Narval  at the  T\'elescopes CFHT and Bernard Lyot, respectively. The most prominent difference between our results and those reported by the authors occur for the star HIP26752, which has the biggest reported error bar. For the spectroscopic binary HIP25258 we fit a deblending function using the Levenberg-Marquardt method with initial values determined by the line centers obtained with FXCOR. Differences with other studies are likely due to observations carried out in different epochs, which is expected to cause variations in the RVs of binary system  components.\\

Heliocentric radial velocites (RVs) for the $\sigma$ Orionis cluster were estimated by  \citet[][see Table 5]{hernandez2014} using the IRAF package RVSAO that  cross-correlates each observed spectrum with a set of synthetic solar metallicity stellar templates from \citet{coelho2005}. Authors reported a list of binary candidates that either  exhibit double peaks in the cross correlation functions or present strong RV variability. In Table \ref{tab:tableTTs} we labeled these objects with “b” and “c”, respectively.  Except for those objects,  the TTs sample can be considered as formed by 
single objects.

\subsection{Rotational velocities} \label{subsec:rotvel}

Projected rotational velocities ($v$sin$i$) for  TTs and HAeBes were obtained applying two methods: 1) analysis of the cross correlation function of the object spectrum against a rotational template \citep{tr1979,hartmann1986} and 2) Fourier Transform (FT)  of selected line profiles in the object spectrum  (Serna et al.2021, in preparation). 

\subsubsection{Cross correlation function analysis} 

In the CCF based method, the $v$sin$i$ of the star is obtained through the calculation of the cross correlation function of each object  against a stellar template of similar spectral type artificially broadened at different velocities. This manufactered broadening was done by using the Python routine \textit{rotbroad} of \textit{PyAstronomy} package with a solar limb-darkening coefficient of $\epsilon=0.6$  \citep{rama1954,hartmann1986}.  For each broadening velocity, we  fitted the peak of the cross correlation function  with a parabola  and measured its full width at half maximum (FWHM) and its S/N through the TDR parameter \citep{tr1979}. Rotational projected velocity is given by the minimum value of the FWHM of the parabola that better fits the peak of the CCF by a calibration function \citep{sacco2008}. This function is computed from a template-template CCF, which is broadened in  $2$ km $s^{-1}$ steps.  The procedure is carried out on the spectral window ($6580${\AA}$<\lambda<6700$ {\AA}),
supressing any emission feature in the observed spectra before the correlation is computed.\\

For the Hectochelle sample (i.e. TTs), we used synthetic templates from the LTE static parallel line-blanketed ATLAS9 model  of \cite{kurucz1979}.  We assumed solar metallicity and a range of surface gravities  4.0 $<$ log g $<$ 5.0 and effective temperatures 3500 K $<$ {\Teff} $<$ 5000 K depending on the spectral type reported by \citet{hernandez2014}. The spectra were degraded to the Hectochelle instrument resolution 
before starting the broadening at steps of $2$ km $s^{-1}$.  As an example, the first panel of Figure 2-a exemplifies the $v$sin$i$ determination for the K1 star 2MASS J05385410-0249297 (upper spectrum) using a template with {\Teff} =4500 K and log g = 4.5 (lower spectrum).  Through the use of the calibration function, the minimum of the CCF at $27$ km $s^{-1}$ leads to  a $v$sin$i$=($31\pm2$) km $s^{-1}$.  Uncertainties in CCF method depend on the errors in the correlation process and are calculated  using the $R$ parameter (see Section \ref{subsec:radvel}). In Col[7] of Table \ref{tab:tableTTs} we report  $v$sin$i$ CCF measurements for TTs. Due to low S/N and uncertainties in the method, almost 30\% of the sample exhibits the lowest measurable rotational velocity of $\sim9$ km $s^{-1}$, which corresponds to the instrumental broadening.   \\

The FIES sample in the same spectral window is characterised by the absence of single lines with good S/N  which adds noise to the cross correlation function.  However, in contrast with other methods,  CCF is able to deal with this fact, since all spectral features are included via a quadratically sum \citep{tr1979}.  For HAeBes, we constructed two calibration functions with the observed templates listed in Table 2 and one with a synthetic spectrum (with spectral type A0; ATLAS9 model of \citep{kurucz1979}). Figure 2-b illustrates the $v$sin$i$ determination for the B9 star HIP26752, the most massive HAeBes in our sample (upper spectrum in the first panel). Original and broadened stellar templates are plotted with thick and thin lines, respectively. In the middle panel, minima in the FWHM of the CCF were identified at 129 km $s^{-1}$ for HD53244, 134 km $s^{-1}$  for HD87737, and 125 km $s^{-1}$  for the synthetic template. Using the calibration functions (dashed lines), we obtain a mean value $v$sin$i$=( 128$\pm$7) km $s^{-1}$ for HIP26752 computed with the three templates. For the particular case of the stellar template HD53244, the resultant CCF (bottom right panel) is slightly higher than the noise ($R = 3$). Despite showing chemical spots leading to rotational and radial velocity variability \citep{briquet2010}, we obtain $v$sin$i$=( 129$\pm$9) km $s^{-1}$  using this star as a template. The fast rotation of HIP26752 that leads to a strong line-blending seems to be the main source of broadening of the correlation peak as pointed out in previous studies \citep{royer2002,diaz2011}. For the rest of HAeBes, we obtain the best results using HD87737 as a rotational template. The $v$sin$i$ CCF values are shown in Col[4] of Table \ref{tab:tablehaebes}.\\

\subsubsection{Fourier transform (FT) analysis} 

Some limitations of the CCF method are: i) the assumption that rotation is the dominant broadening process of the photospheric lines \citep{wilson1969}, ii) a strong dependence on the calibration function and iii) variations in $v$sin$i$ depending on the spectral window \citep{hartmann1986}. These factors are avoided thanks to methods based on Fourier transform (FT) of selected line profiles. The method relies on the fact that  the FT of the observed line has zeros whose  location is related to the $v$sin$i$ of the star.  Specifically, if we denote with $\nu_1$ the first zero of the FT of a line profile at $\lambda_0$,  the resultant rotational projected velocity is given by $v$sin$i=c\Delta\lambda/\lambda_0$ where $\Delta\lambda=\nu_1/q_1$, here $\nu_1$ the first zero of the FT and $q_1$ a polynomial function that depends on the limb darkening \citep{carroll1933}.  Under the assumption that TTs and HAeBes share same limb darkening  physics, we adopted $\epsilon=0.6$ and subsequently $q_1=0.660$ in all calculations.  \\

We applied this method to our studied sample as follows. For TTs we focused mainly in the Li6707.74/89{\AA} doublet which has good S/N in our Hectochelle spectra.  For a few cases  where the doublet is not present, or it is contaminated by merging of spectral orders, we use the stellar spectral features  FeI6575{\AA}\ , 6696{\AA}. We report rotational projected velocities for  72 objects using FT method labeled as $v$sin$i^{Fourier}$ in  Table \ref{tab:tableTTs}.  For HAeBes we applied FT to  MgII4481{\AA}, FeI4489{\AA} and FeI4226{\AA} line profiles which have been tested as excellent rotational indicators in previous studies of stellar rotation. Values appearing in Col[4] of Table \ref{tab:tablehaebes} correspond to the average of the $v$sin$i^{Fourier}$ values obtained from all lines.  Comparison with the CCF based method is  indicated with open squares in Figure \ref{fig:f2}. Absolute differences between Fourier and CCF methods  remain below $5$ km $s^{-1}$ in the worst of the cases and slightly larger values for HAeBes. In view of the strong dependence of CCF method with the calibration function, we find reasonable  to keep the with $v$sin$i^{Fourier}$  estimates, instead  of the upper values obtained from $v$sin$i^{CFF}$ in the TTs sample.    
 
\subsubsection{Comparison with previous studies}

We compared the rotational velocities obtained through the CCF technique  with those previously published. In the case of TTs, \cite{sacco2008} conducted a similar study in $\sigma$ Orionis based on FLAMES/VLT spectroscopic observations (R = 17000),
using a similar technique for $v$sin$i$ determinations. The authors obtain  upper limits of $17$ km $s^{-1}$ for the majority of the objects with the exception of J05382774-0243009 and J05383431-0235000 whose $v$sin$i$ of $23.7$ and $31.7$ km $s^{-1}$ are in agreement with our measurements. More recently,  \citet{kounkel2018} and \citet{kounkel2019}  analysed spectroscopic data from APOGEE-2 ($R = 22500$) and reported  $v$sin$i$ for 45 of our objects. Comparison with our data is indicated by gray squares in the left panel of Figure \ref{fig:f2}. Our  $v$sin$i$ values are in agreement with those reported by other authors for velocities above resolution limit of APOGEE-2 i.e. 13 km $s^{-1}$. Concerning HAeBes, we compared our $v$sin$i$ values obtained with CCF method with those obtained from the spectropolarimetric analysis conducted by \cite{alecian2013}. Symbols in black in the right panel of Figure \ref{fig:f2} indicate absolute observed differences below $8$ km $s^{-1}$.

\section{Rotation of TTs and Herbig Ae/Be stars} \label{sec:swpa}

In Figure \ref{fig:all} we show the variation of the rotation rate $v$sin$i$ with  stellar mass obtained through the methodology described in Sections \ref{subsec:rotvel} and \ref{subsec:masa}, respectively. We recall that the majority of the objects are  reliable members with ages between 3 and 10 Myr and therefore  this distribution represents well the stellar rotation within this age interval. While the age normally adopted for the $\sigma$ Orionis cluster is 2-4 Myr e.g.\citep{zapatero2002,penaramirez2012}, the ages of Orion OB1a and OB1bc have  upper limits of 10 Myr. Nonetheless, high dispersion in the rotation rates for stars younger than 10 Myr is a typical feature among TTs in other young clusters such as ONC  
\citep[$\sim$ 1Myr;][]{herbst2002}, Taurus,  $\rho$-Oph \citep[$\sim$ 1-2 Myr;][]{rebull2018}, 
and the $\sigma$ Orionis cluster \citep{cody2010}. In fact for our sample of TTs we obtain  $<v$sin$i>^{TTs}=(21\pm9)$ km $s^{-1}$. \\

We separate the TTs into accretors and non-accretors, according to the classification 
 by \cite{hernandez2014} based on the $H\alpha \,  10\%$ line. We find that  53\% correspond to accretors,  27\% are non-accretors,  8\% are binaries candidates and 12\% lack of accretion information.  Rotational velocities are slightly  larger for non-accretors ($<v$sin$i>^{non-acc}=(24\pm9)$ km $s^{-1}$) in comparison with accretors  ($<v$sin$i>^{acc}=(19\pm8)$ km $s^{-1}$) as expected from MA and confirmed in stellar associations ($<$10 Myr) by \cite{jay2006}.  Accreting stars rotate, on average, slower than non-accreting stars.  This 
agrees with the scenario in which the stellar rotation in CTTS is affected by the disc braking phenomena \citep{bouvier2013}. \\

Concerning HAeBes, they exhibit rotation rates significantly larger than  those in TTs. We report  $<v$sin$i>^{HAeBes}=(115\pm9)$ km $s^{-1}$ confirming  an increase in $v$sin$i$ with  stellar mass from TTs to HAeBes by a factor of $\sim$5. We indicated our six objects with squares in  Figure \ref{fig:all}. While single HAeBes remain roughly constant above $\sim100$ km $s^{-1}$, the binary system HIP25258, is placed on the bottom of the diagram (open squares). Complementary data of HAeBes of  \cite{alecian2013} and \cite{fairlamb2015}  are also included and shown with filled circles in grey. The binary candidates are indicated by empty circles with upper limit values indicated by crosses.  The two slowest rotators HD190073 and BD-051253 are accreting, being active HAeBes.  While the former is a $3$ $M_{\odot}$ active HAeBe \citep{manoj2006}, the later is a B9 star with  an accretion rate
of log $\dot{M}_a=-5.34$ $M_{\odot}/yr$ \citep{fairlamb2015}.

\section{ The interplay of accretion and winds in HAeBes}\label{subsec:profile}

In order to confirm whether HAeBes are accreting or not, we searched for excess fluxes in the Balmer discontinuity relative to a stellar template of similar spectral type. Assuming a magnetospheric accretion framework, this excess is related to the accretion rate of protoplanetary discs \citep[e.g.,][]{rigliaco2012,donehew2011,mendigutia2013}. In Figure \ref{fig:f9} we show the  normalized spectrum of each star to the stellar template in the interval that includes the Balmer jump. Despite of the contamination due to emission lines, excess emission in the Balmer discontinuity is observed in all objects. Under the hypothesis of magnetically controlled accretion, this behavior suggests that active accretion processes are present in the HAeBes sample \citep{cauley2015,villebrun2019}. \\

Regarding the presence of winds, we examined the morphology of the residual profiles of the FIES spectra searching for mass loss features in the form of jets and blue-shifted forbidden emission lines such as [OI]$\lambda$6300{\AA}. Residual profiles were obtained after subtracting the photospheric contribution  using the spectrum of a standard star, rotationally broadened to the $v$sin$i$ of the target. Panels (a), (b) and (c) of Figures \ref{fig:buena} to \ref{fig:f8}  respectively show such profiles for the $H{\alpha}$, $H{\beta}$ and $H{\gamma}$ Balmer lines. The forbidden line [OI]$\lambda$6300{\AA} is shown in Panel (d) whereas in Panel (e) we show  HeI$\lambda$5876{\AA} whose physical origin comes from magnetically controlled accretion  \citep{beristain2001}. Finally, the CaII$\lambda$3933{\AA}, recognized as an  indicator of chromospheric activity in TTs,  is shown in panel (f). We note that all HAeBes exhibit some residual level of emission in this line, relative to the photospheric contribution (dotted lines). These six profiles were classified  in groups as follows:  Double-peak (DP), Inverse P Cygni (IPC), P Cygni  (PC), (E) in emission, (A) in absorption and (F) flat. It is apparent that  Balmer lines  show complex profiles with red, central and blue-shifted absorption features that reveal presence of  accretion phenomena of even active chromospheres with inflows and outflows of matter. \\  
 
 In Figure \ref{fig:buena} we show line profiles for HIP26955, an object with a clear PC profile in the Balmer lines, with a blue-shifted absorption feature dipping below the continuum at $\sim$-200 km $s^{-1}$ and with the edge extending up to $\sim$-300 km $s^{-1}$, a clear evidence of a stellar wind. This object also shows emission in [OI]$\lambda$6300{\AA}  with a broad component  centered at  $\sim$30 km $s^{-1}$ and  extending  up to 60 km $s^{-1}$.  Although a similar behaviour in less degree is exhibited by   HIP26752 (Figure \ref{fig:hip26752}), the other stars show a very narrow residual emission  centred  around $-30$ km $s^{-1}$ and with a FWHM of $\sim$5 km $s^{-1}$ which is due to night sky line contamination.   The [OI]6300{\AA}  emission line originates in low density regions, in the outer parts of stellar winds, and its broadening is  associated  to the terminal velocity of the stellar wind \citep{bohm1994, hartigan1995}.  In addition to showing forbidden line emission, the HAeBes HIP26955 and HIP26752  show the largest dereddened  WISE colour indexes  $[3.4 - 22]_0\mu$m among studied HAeBes as confirmed by its parameters in Table \ref{tab:tablehaebes}. \\ 
  
In TTs, forbidden line emission correlates with infrared excesses, which is interpreted as winds that are powered in some way by the stellar accretion.  High-spatial resolution studies are required for exploring the physics of the wind-launching mechanism in these systems \citep{hone2017}. Assuming that the stellar accretion powers stellar winds, theoretical models suggest that TTs with mass loss rates of one tenth of the accretion rate can lose enough angular momentum to keep the stellar rotation locked during the first 3 Myr. \citep{matt2005,cranmer2008}. This is supported by the observed morphology of the majority of Balmer emission lines in high spectral resolution, with the presence of simultaneous red-shifted and blue-shifted absorption features that are interpreted as accretion and winds events, respectively. In contrast, census of both red and blue-shifted absorption features in large samples of HAeBes confirm lower occurrences in comparison to TTs as well as differences in the accretion mechanism between Herbig Aes and Bes, suggesting that innermost environments of HAeBes could not be a scaled version of those of the TTs \citep{cauley2015}. An eventual transition from the magnetospheric accretion/ejection paradigm into other mechanisms such as the boundary layer scenario is thus expected, but more detailed studies are required.\\

\section{ Synthetic $v$sin$i$-mass distributions }

We derive a set of relations in the $v$sin$i$-mass diagram for different disc lifetimes and magnetic field strengths to study general trends between rotation and stellar parameters such as accretion, magnetic field, and the presence of a disc on a spin evolution model.
Rather than attempting to explain all phenomena involved in the rotational evolution of TTs and HAeBes, the main goal is to address the question of to what extent the current picture of angular momentum evolution in TTs can be extrapolated toward HAeBes. The source code and all required files are of public domain\footnote{\url{https://github.com/gpinzon/REFUGEE} }.

\subsection{Model assumptions}

The model is an extension toward lower and higher masses of the one described by \citet{matt2012} and it is used to compute the rotational evolution of a solar mass star, magnetically linked to a surrounding accretion disc   during the Hayashi track.  We compute the rotational evolution, for a wide range of  masses,  from the birth line to the adopted age for $\sigma$ Orionis (3 Myr). We present calculations of the spin rate of protostellar masses between 0.1 and 7 $M_{\odot}$ at 1, 3, 5 and 10 Myr, considering a range of disc timelifes between 0.08 and 3.0 Myr and magnetic field strengths  in the interval 500 to 3000 \textit{G},  representative of TTs and HAeBes. The model relies on the following assumptions: \\

\textbf{H1} Solid body rotation : To model the rotational evolution, we assume  uniform internal rotation, although the stellar interior is described by non-rotating evolutionary stellar models.  The internal structure of stars is obtained from the grid of 27 PMS mass tracks of \cite{siess2000}  for solar metallicity (Z=0.02) spanning over 0.1 to 7.0$M_{\odot}$. We find reasonable to assume that TTs are well described by the interval {\boldmath $0.1\le M_*\le1.5M_{\odot}$ whereas  HAeBes fit the condition $M_*>1.5M_{\odot}$ \citep{hillenbrand1992}.} \\

\textbf{H2} Disc-locking:   This effect arises from the magnetic interaction between the star and the surrounding gaseous disc. This stage lasts a few Myr \citep{bouvier2013} and depends on both,  magnetic coupling strength to the disc and the opening of magnetic field lines due to differential rotation \citep{uzdensky2002}.  We used  the same  assumptions described in \cite{matt2005} and \cite{matt2010} for the calculation of this magnetic torque that are summarized as follows: (1) a critical twisting $\gamma_c=1$ which  represents comparable azimuthal and vertical magnetic field components within the disc in order for the dipolar field lines to remain closed  and (2) a magnetic diffusivity parameter  $\beta=10^{-2}$ which describes strong coupling between the star and disc \citep{rosen2012}. The disc-locking durability is given by the disc lifetime $\tau_D$ which is a free parameter of the model. We consider the following cases: $\tau_D$=$0.08$, $0.5$, $0.7$, $1.0$, $2.0$ and $3.0$ Myr. \\

\textbf{H3} Stellar winds:  In TTs, powerful winds arising from open field regions, i.e. accretion powered stellar winds (APSWs) have been shown to be  the primary or one of the most important agent for removing angular momentum from the star \citep{hartmann1982, romanova2005, matt2012}.  It is
 assumed that in these stars a fraction $\chi$ of the energy released during the accretion process  is dissipated close to the star surface and transferred to the stellar wind. The stellar wind torque $T_w$ in the APSWs scenario is  given by:

\begin{equation}
 T_{w}=-\dot{M}_w\Omega_*r_{A}^2,
 \end{equation}
 
 where  $\dot{M}_w=\chi\dot{M}_a$ is the mass loss rate in the wind,  $\dot{M}_a$ is the stellar accretion rate, $\Omega_*=v_*R_*$ is the angular velocity of the star and $r_A$ is  the location where the wind speed equals that of magnetic Alfv\'en waves. These radii were computed based on solutions for two-dimensional axisymmetric solar-like stellar winds \citep{matt2008}. The mass-loss-weighted average of the Alfv\'en radius in the multi-dimensional flow is thus computed through  the relation:

 \begin{equation}
 \frac{r_A}{R_*}=K(\frac{B_*^2R_*^2}{\dot{M}_w v_{esc}})^m.
 \end{equation}
 
   Here $B_*$ is the stellar magnetic field, $R_*$ the radius of the star, $\dot{M}_w$ the wind mass loss  rate and $v_{esc}$ the escape velocity.   In concordance with the main purpose, which is an extension of the model toward higher masses maintaining same physics, we fixed $K$ and $m$ in all our simulations. Following \citet{matt2008} and \citet{matt2012} we adopted $K=2.11$ and $m=0.223$ which are in agreement with recent 2.5D magnetohydrodynamic simulations of \citet{pantolmos2021}.     Concerning the wind mass loss, it is computed at any instant of time through  $\dot{M}_w=0.1\times\dot{M}_a$ i.e. $\chi=0.1$ \citep{hartmann1989,matt2005,cabrit2007,ahuir2020}.  Finally, we recall that $T_w$ is null for $t>\tau_D$.\\

\textbf{H4} Changes in the accretion rate : We assume that accretion depends  on both mass  and time as follows: 
\begin{equation}
\dot{M}_a(M_*,t)=\dot{M}_{a,0}(M_*)e^{-t/\tau_a}, \label{uich}
\end{equation}

 where  $\dot{M}_{a,0}$ is the accretion rate at birthline  and $\tau_a$ is the characteristic timescale for the temporal decay. The decay of accretion cannot be explained entirely through an empirical relationship with age, however  observations in open clusters confirm a decay as $t^{-k}$ with $k$ between $-1.6$ and $-1.2$ \citep{hartmann1998,sicilia2010,manzo2020}. For the exponential decay in equation \ref{uich} we find plausible to adopt $\tau_a=8$ Myr in all simulations,  which is an intermediate value between the average disc lifetime of 3-5 Myr used by \citep{gallet2015} and maximum lifetimes of 10-20 Myr measured by \cite{bell2013}.  \\
 
   The accretion rates for TTs and HAeBes, determined from spectroscopic observations, correlate with the mass of the star through a power-law  $\dot{M}_a\propto M_*^\alpha$ with $\alpha$ between $1.5$ and $3.1$ \citep{muzerolle2004,manara2015}. Under the assumption that this correlation is maintained at the birthline as well, we are able to compute initial accretion rates for a wide range of masses. We conducted a compilation of accretion rates at 3 Myr as shown in Figure \ref{fig:f10a}.   For TTs we used members studied by  \citep{rigliaco2012} and \citep{mauco2016}, whereas for HAeBes we used data from  \citep{alecian2013} and \citep{fairlamb2015}.  By considering the sample as a whole, the best fit is reached with  $\alpha=(2.51\pm0.20)$  in agreement with the exponents obtained  in other star forming regions such as Taurus \citep{calvet2004} and Ophiucus \citep{natta2006}.  Although a more steeper correlation ($4.6<\alpha<5.2$) between accretion rate has been reported in HAeBes \citep{mendigutia2012} with notable differences between HAes and HBes.  The scatter in Figure \ref{fig:f10a}  remains constant at about two orders of magnitude throughout and is explained by variability, errors in mass estimation from stellar models and bias attributed to sample selection  \citep{hartmann2016}.   \\

\textbf{H5} Magnetic field strength :  It is assumed that  the star has a uniform co-rotating dipolar magnetic field with strength $B_*$ anchored to its surface. This stellar field connects to an 
 extended disc region, reaching beyond the disc co-rotation radius depending of its strength which, in turns, depends on the spectral type.  For  TTs with masses above the convective limit ($M_*\gtrsim0.3M_{\odot}$), the assumption that  magnetic field  forms via a solar-type dynamo $\alpha-\Omega$ is consistent with the observed strengths in the range $1<B_*<5$ kG \citep{vidotto2014}.  In addition, magnetic fluxes of stars with masses between $0.3$ and $2.0M_{\odot}$, scale up with the rotation rate until the saturation which is supported by the stellar dynamo theory.  In fully convective stars ($M_*\lesssim0.3M_{\odot}$), the absence of a interface dividing radiation from convection prevents the generation of magnetic fields via a solar-type dynamo. However, theoretical models suggest that small-scale fields may be amplified and transformed into non-axisymmetric large-scale fields  under the action of differential rotation \citep{dobler2006,browning2008}. Very low mass TTs ($M_*\lesssim0.2M_{\odot}$) display a variety of strengths and topologies being the large-scale fields predominantly poloidal and axisymmetric, with strengths in the order of a few kG as confirmed through the analysis of their highly polarized rotationally modulated radio emission  \citep{berger2006}. \\

On the other hand, magnetic fields in HAeBes are scarce with less than 10\% of large samples hosting large-scale dipolar fields stronger than $0.3$ kG \citep{wade2007,alecian2013}.  We note that this statistic is based on measurements with uncertainties too large that prevent detections of the order of a few tens to a few hundred $G$.  The study by \cite{hubrig2015} suggests that the low detection rate of magnetic fields in HAeBes can be plausibly explained by the limited sensitivity of the published measurements and by the  weakness of the magnetic fields in these stars. Regardless of these factors,  the large-scale fields among the few magnetic HAeBes,  have  similar strengths to those displayed by the slow rotating $Ap/Bp$ stars on the main sequence \citep{auriere2007,kb2006}. The presence of magnetic fields in HAeBes and their disappearance at evolutionary stages closer to the main sequence (see \cite{hubrig2009a,hubrig2015}) indicate that these fields are probably remmants of the  magnetic fields generated by dynamos during the convective phases at early PMS stages. In this context, HAeBes with strong kG fields can be considered  progenitors of Ap/Bp stars  \citep{ferrario2009}. For this work, we find it reasonable to assume that fields in HAeBes are already present at the birth line and survive the pre-main sequence all along. The rotational models presented here, were computed separately for the four constant strength values of $B_*=0.5$, $1.0$, $2.0$ and $3.0$ kG.

\subsection{Numerical Method.}  \label{subsec:numericalmethod}

The evolution of angular velocity $\Omega_*$ is computed as follows:

 \begin{equation}
 \frac{d\Omega_*}{dt}=\frac{T_*}{I_*}-\Omega_*[\frac{\dot{M}_a}{M_*}(1-\chi) + \frac{2}{R_*}\frac{dR_*}{dt}], \label{one}
 \end{equation}
 
 where $I_*$ is the stellar moment of inertia of the star, $\chi=0.1$ \citep{cabrit2007,matt2012} is the fraction of the accretion that goes into the wind and $T_*$ is the net torque. We recall that $T_*$ has three contributions coming from accretion, stellar winds and star-disc interaction. This total torque  is artificially set up to zero for $t\ge \tau_D$ where $\tau_D$ is the gas-disc lifetime.  We used a family of disc lifespans ranging from long-lived discs ($\tau_D=3$Myr) up to discless stars ($\tau_D=0.08$Myr). 
 
 \subsubsection{Initial conditions}
 In absence of information about initial angular velocity, we fixed it at the beginning of the PMS to  one-third of the break-up limit or critic velocity $v_{c}$ in all cases. This assumption is supported  by the fact that under sufficiently high field strengths and accretion rates, the stellar rotation quickly reaches an equilibrium in which $\Omega_*$ is independent of initial rotation \citep{colliercameron1995}. Regarding initial stellar mass and radius,  they were obtained from interpolation based on evolutionary models of \cite{siess2000}.\\
 
 The initial accretion rate values, were computed assuming $\dot{M}_a = M_*^{\alpha}$ with $\alpha=(2.51\pm0.20)$. We integrate the equation \ref{uich} inward on time in order to get $\dot{M}_{a,0}$ for each mass. This results in initial accretion rates in TTs of $\dot{M}_a=5\times 10^{-9}M_{\odot}/yr$  that lead to  initial disc masses of the order of $M_D=0.03M_{\odot}$, compatible with disc masses of $\sim$0.02$M_{\odot}$ estimated from cold dust emission \citep{hartmann1998}.  For HAeBes we obtain $\dot{M}_a=5\times 10^{-7}M_{\odot}/yr$ and thus initial disc masses with a median of $M_D=0.65M_{\odot}$, consistent with values reported by \citep{mendigutia2012}.  Despite being low, the derived values reflect the observed trend with mass displaying high dispersion and therefore might be considered just representative values.   \\

 We integrate the  equations (\ref{uich}) and (\ref{one}) using the fourth-order Runge-Kutta scheme of \cite{press2007} with adaptive step size. At each time-step we calculate the next one by requiring that none of the variables change by more than 1\% per step. This scheme requires just a few hundred timesteps to complete the computation from  birthline $\tau_0=0.08$Myr up to $1$, $3$, $5$ or $10$ Myr. 
  In the computation, the $I_*$ and $R_*$ values from \cite{siess2000} models are interpolated using a third degree polynomial.

\section{Results}

We first consider two rotational histories during the first 3 Myr after birth-line, with the same initial conditions and disc timelife ($\tau_D=3.0$ Myr) but varying the magnetic field strength.  Figure  \ref{fig:f12}
 shows the evolution of the stellar radius $R_*$, Alfv\'en radius $r_A$ and equatorial velocity $v_*$  for 
$B_*=0.5$ kG (panels (a) and (b)) and $B_*=3.0$ (panels (c) and (d)). Models for distinct masses are indicated with thick  lines of different types and colors.  HAeBes are represented by lines in blue, green and red whereas TTs by lines in black with distinct line types. We recall that rotation in TTs and very likely in HAeBes tends toward a rotational equilibrium in which the total torque on the star would be zero.  From Figure \ref{fig:f12} it is clear that in both  groups of stars,  the spin rate increases substantially  with time at the beginning of simulations, due to accretion and the rapid contraction of the star, both adding angular momentum. We note that the Alfv\'en radius in the stellar wind for the model of 5 $M_{\odot}$ and for the case of low field strength, coincides  with the stellar radius  during the first $\sim6\times10^5$ years. Considering that this model reaches the main sequence at 3 Myr, that time represents 20\% of its PMS lifetime.  On the other hand, field strengths in the range of kG enable efficient angular momentum loss since the beginning of evolution. From Figure \ref{fig:f12} it is evident that higher fields produce larger $r_A$ and thus more effective spin-down torque.  We find that for the case of 3 kG, HAeBes undergo a spin down by a factor of almost 3 at the end of simulations.\\

The expected stellar spin up resulting from an early absence of a disc  ($\tau_D=1$ Myr) is indicated in Figures 13-b and 13-d  with thin lines of same types and colors.  Results are in line with a  disc locking scenario in which, once the gas within the disc dissipates due to stellar accretion and disc photo-evaporation phenomena, the contraction proceeds increasing rapidly the stellar rotation \citep{sicilia2010,bouvier2013}.   For both, TTs and HAeBes, the separation between rotational  tracks computed  with $\tau_D=1$ and $3$ Myr results significantly larger for $B_*=3$ kG as illustrated in panels (c) and (d).    \\

 Since equilibrium spin rate is reached by 3 Myr, roughly the  mean age of our sample,  we  computed a set of synthetic mass-$v$sin$i$ relations at  fixed ages  of  $1$, $3$, $5$ and $10$ Myr. These time-snapshots of stellar rotation were calculated for distinct stellar magnetic field strengths $B_*$ and disc lifetimes $\tau_D$ as shown in Figures \ref{fig:f12a} to \ref{fig:f12d}. We consider six disc timelifes  ranging from $0.08$ to $3.0$ Myr and four magnetic field strengths of $0.5$, $1$, $2$ and $3$ kG.  For each pair  ($\tau_D$, $B_*$) we obtain a rotational track indicated with a dashed curve in the mass-$v$sin$i$ diagrams. In all cases, tracks lie below the critical velocity $v_c$ (dotted line) with a pronounced dip at $4.0 \, M_{\odot}$, consequence of the shell burning of deuterium that swells up the star significantly and thus induces a sudden spin down \citep{hosokawa2009}. We divide the evolution in this mass region in a swelling phase for $2M_{\odot}\lesssim M_*\lesssim4M_{\odot}$ followed by a phase of rapid gravitational contraction for $ M_*\gtrsim4M_{\odot}$.  From Figures \ref{fig:f12a} and \ref{fig:f12b} it is clear that field strengths of hundred of G are unable to predict the existence of slowly rotating TTs and HAeBes. It is  also clear the poor  impact of distinct $\tau_D$'s, especially  at $1$ Myr. It can be seen that a large number of TTs with active accretion are located in a region 
 not covered by the models. In Figures \ref{fig:f12c} and  \ref{fig:f12d} higher field strengths lead to  larger splitting of tracks corresponding to distinct disc timelifes confirming that the presence of a disc is compatible with fields strengths of the order of kG as expected for TTs \citep{muzerolle2004,gallet2015}.  In  particular the set of rotational tracks associated to the pairs ($10$ Myr, $2$ kG) and ($\tau_D\ge3$ Myr, $3$ kG) match well the range of velocities and masses covered by our sample.  

\subsection{Angular momentum evolution in Herbig Ae/Be }
 
In order to quantify differences between  the data and  the models, we used the specific angular momenta $J$sin$i/M_*$ instead of $v$sin$i$.  In this section we first describe the systematic trends with mass and thus compare with specific angular momenta obtained using our rotational model.  Figure \ref{fig:f13} represents the  angular momenta for all sample of TTs (triangles) and HAeBes (rectangles) obtained through $J$sin$i/M_*=k^2R_*v$sin$i$  where the gyration radii are given by \cite{siess2000} . We see a  gradual increasing  of angular momenta with mass in the interval 0.1-4.0 $M_{\odot}$, with a mean  value  of $5\times10^{17}$ $cm^2$ $s^{-1}$ for HAeBes.  As a reference, the dashed grey line represents the empirical relationship $<J/M_*>\propto M_*^{1.02}$ \citep{kawaler1987}, valid for mature main-sequence stars and   constructed on the basis of stars rotating as solid bodies to a fixed fraction  of their critic value.  While  projection effects  contribute  less than 15\%  to the scatter \citep{wolff2004}, changes in the magnetic fields and fast disc dissipation could be playing  an important role. \\

In addition, in Figure  \ref{fig:f13}, the angular momentum tracks were computed for the case $\tau_D=3.0$ Myr with the exception of  the  dotted line which represents discless stars. Although the dispersion is large,  the data scatter is fitted reasonably well by
the tracks corresponding to $B_*=2.0$ and  $3.0$ kG.  Table \ref{tab:ks} shows the Kolmogorov-Smirnov (KS) statistic and  p-value or probability that observed and expected values  are drawn from the same distribution. For TTs with $B_*\le 1$ kG we can reject the null hypothesis since the p-value is neglictible. However, for $2$ and $3$ kG,  the hypothesis can not  be rejected since p-values are  71\% and 10\%, respectively. The best match of TTs with the model occurs for $B_*=2.0$ kG which has the smallest KS statistic and the largest p-value. Concerning HAeBes, the highest p-value of 63\% is obtained for $B_*=2.0$ kG as well.  The other strengths lead to p-values below 24\% and therefore we can reject the null hypothesis for them.  For the particular case of $B_*=2.0$ kG indicated with a thick dot-dashed line, we compute their separation  from the main-sequence finding that on average,  TTs {\boldmath ($<1.5M_{\odot}$)}  have  larger specific angular momenta by a factor of $5/2$.  This dramatic  
 increase in the angular momentum between $2$ and $3$ $M_{\odot}$ is due to structural changes related to both, deuterium burning and the  reaching to the main sequence. In this region, $<J/M>$ is larger than in the main-sequence by a factor of $3.2$. \\

Fields strengths of a few kG  are in line with measurements of circular polarization of Ap/Bp main sequence stars \citep{alecian2013}. In Figure \ref{fig:f13} we indicate with pentagons the specific angular momentum computed for the sample of $23$ Ap/Bp stars studied by \cite{auriere2007}.  Masses and ages were computed using evolutionary  MIST models by \citep{dotter2016},  via  effective temperatures and luminosities reported by the authors and the methodology described in Section \ref{subsec:masa}.  They find longitudinal components larger than a few tens of $G$ and use this information to infer the dipolar component of the field.  The size of the pentagons  in Figure \ref{fig:f13} is proportional to this dipolar component whose minimum and maximum values are $0.10$ and $8.9$ kG, respectively.  Authors report a plateu at about $1$ kG falling off to larger and smaller fields. Two main groups in the form of bimodality are observed. A slow group of Ap/Bp with roughly high dipoles and  $<J$sin$i/M_*>$=$6.3\times10^{15}$ $cm^2$ $s^{-1}$. A second group  with weak dipolar component with larger $J/M$'s   by a factor of $\sim$6.  \\
 
How to connect  these results with angular momentum in HAeBes is not clear yet,  in particular since Ap/Bp stars are found mostly in clusters older than $10^8$ yr \citep{abt1979}.  Several works, have pointed out that these stars do not experience substantial magnetic braking during their life on the main sequence \citep{north1998,kb2006}. However, this conclusion depends strongly on the origin and evolution of magnetic fields, a subject that still is  under debate.  Basically, there are three  scenarios: 1) the magnetic field  appears once the star has spent a considerable fraction of its existence on the main-sequence  \citep{hubrig2000}, 2) magnetic fields in Ap/Bp stars are present at the birth-line already \citep{moss1989,kb2006}.  Studies based on analysis of chemical anomalies suggest  that magnetic fields are shaped during the first Myr and do not have considerable changes since  \citep{gomez1998,pohnl2005,wade2007}. This is in agreement with theoretical predictions for the ohmic decay of the field  inside stellar interior which is  non-neglictible,  only in scales of Gyr, much longer than the main sequence lifetime for Ap/Bp stars \citep{moss1984}.  3) The magnetic field appears during PMS stage, as a consequece of a merging between two low mass stars \citep{ferrario2009}. This merging event occurs when one of the two objects at least, has arrived at the end of its Henvey track in the HR diagram \citep{iben1965}. The resultant merged object has stronger  differential rotation  and  therefore a large-scale dynamo field. \\

In  Figure \ref{fig:f13} we have included complementary data of HAeBes from the compilation by \cite{alecian2013}. Single stars  are indicated with rectangles in gray whereas binaries are indicated with open symbols. The two slowly rotating objects located in the Ap/Bp region are  HD190073 (spT A1) and BD-06 1253 (B9). The former has  marginal reported detections of magnetic field, although  it  displays $\lambda$ Boo-like chemical peculiarities \citep{castelli2020}.  Their low rotation is explained due to inclination  effects  \citep{jarvinen2019}.  The star  BD-06 1253 exhibits weak Ap/Bp peculiaties on its surface composition.  \cite{reipurth2013} have proposed that this star belongs to a quadruple system formed by a Herbig Be, which in turns is a spectroscopic binary \citep{leinert1997}.  In addition, BD-06 1253 is a possible source of outflows  observed in radio frequencies  \citep{rodriguez2016}.   However magnetic field measurements of this object are uncertain since no periodicity was found in the behavior of the most prominent emission lines \cite{alecian2009}. Therefore, it is quite possible that the chemically peculiar component with the detected dipolar magnetic field is not a HAeBe, but already a star at an advanced age, probably on the main-sequence. On the other hand, the Herbig Be status of the primary component is merely based on the appearance of emission in the above mentioned lines belonging to the TTauri component. \\

  Are these sources descendents of HAeBes ?.  
  Under the assumption that magnetospheric accretion in HAeBes is just a scaled 
  version of TTs, we find this probable.   If we assume that at 3 Myr, magnetic fields in HAeBes have already formed via merging \citep{ferrario2009},  then the stellar angular momentum can be transferred outward through outflows in the form of winds and magnetic interaction with the disc if any.  Wind torques could be applied onto the star up to the main-sequence and beyond.  By analysing the computed ages for the Ap/Bp sample,    we find that the majority are younger than 10 Myr (blue pentagons), suggesting that the merging scenario seems compatible with the fact that the loss of angular momentum must occurr very soon during the PMS phase \citep{north1998}.  We  find that  around $\sim4.6\times10^{17}$ $cm^2$ $s^{-1}$ of specific angular momentum must be lost in a few hundred Myr, from  the HAeBes  phase to Ap/Bp.   HAeBes have larger $J/M$'s than Ap/Bp,  by a  factor of 12 for the faster group, and almost 80 for the
slowest rotating but highly magnetic Ap/Bp one.\\
 
 Compared with normal (non-magnetic) stars on the upper main-sequence, Ap/Bp are slow rotators. From Figure \ref{fig:f13} we can easily see that specific angular momentum for the faster group of Ap/Bp stars is about 25\% of that for stars of similar spectral type \citep{kawaler1987} and on the main-sequence, and about 10\%  in the case of highly magnetic Ap~/Bp stars. 
 
\section{Summary and Conclusions} \label{sec:conclusions}

Based on a sample of young stellar objects belonging to the molecular complex of star forming regions in Orion, we computed $v$sin$i$ values for  73 TTs and 6 HAeBes  using two independent methods through a careful analysis of high resolution spectra obtained with FIES and Hectochelle instruments.  Radial velocities, visual extinction values, masses, radii and ages were also computed. Radial and rotational projected velocities obtained from FIES spectra are in agreement with previous studies 
\citep{alecian2013, fairlamb2015}. For our HAeBes sample we obtain a median of $<v$sin$i>=(115\pm9)$ km $s^{-1}$.  Rotational velocities for the $\sigma$ Orionis cluster using Hectochelle are independent of the implemented method (CCF and Fourier) under the uncertainties of each one. \\

We visually inspect the residual line profiles of HAeBes finding evidences of accretion and winds, in particular  HIP26955 is a star that displays PC profiles in all Balmer lines, significant [OI]$\lambda$6300{\AA} emission and large infrared [3.4-2.2]$\mu$m WISE excess. While most prominent emission lines in all HAeBes show complex profiles with red, central and blue-shifted absorption features and all of them exhibit Balmer excesses, only HIP26955   exhibits forbidden line emision in [OI]. These characteristics make it an excellent candidate for testing the magnetospheric accretion model.  \\ 

With the aid of a  rotational model, we investigated the trends  in the $v$sin$i$ $vs$ mass diagram for masses between 0.1 and 7.0$M_{\odot}$ when changes in accretion rates, magnetic field and disc durability are included. The model includes a variable lifetime for the gas in the disc that marks the end of any torque acting on the star.  It is assumed that accreting stars  rotate as solid bodies and that  they regulate their angular momentum through stellar winds powered by accretion. We adopted a uniform stellar dipolar field with constant strength $B_*$. For intermediate-mass stars we suppose that this field was originated during the fully convective phase and has survived since \citep{wade2007}.  \\

By assuming that TTs and HAeBes are surrounded by gaseous discs during the first 3 Myr after birth-line, the best fit to the data was obtained for $B_*=2.0$ kG. For this particular case, we obtained a set of relationships between stellar angular momentum and mass for different disc  lifetimes ranging from discless stars to stars with long-lived discs. We used these relationships, together with the Kawaler law, to estimate the amount of especific angular momentum that must be  lost during 
 the contraction  
 towards the main sequence. In one hand, our results 
 predict that HAeBes stars must 
 lose angular momentum by a factor 
 of $ \sim$3.2, equivalent to an amount of specific angular momentum equal to $\sim3.2\times10^{17}$ $cm^2$ $s^{-1}$. On the other hand,  $<J/M>$ in TTs is larger by a factor of 5/2 than in the main-sequence. \\

We complemented our $<J/M>$ values for HAeBes with the 
 sample from \citet{alecian2013} in order to compare  our models with observed data of a particular sample of Ap/Bp stars analysed by \cite{auriere2007}. We find that 
specific angular momentum must be lost by a factor between 12 and 80 from HAeBes to Ap/Bp stars, depending on the intensity of the dipolar field.\\

Although detailed phenomena of TTs and HAeBes, such as stellar and disc inclination,  disc photoevaporation,  complex topologies of magnetosheperes,  among other factors are not considered,  
the model presented here, based on  simple assumptions is extremely useful for testing the impact of rotation on distinct physical stellar parameters  during the evolution of young stellar objects over a wide range of spectral types.  However, the results obtained for the angular momentum in HAeBes do not explain the low rotation of Ap/Bp stars.    \\

\vspace{1.5em}
We thank the anonymous referee for providing many insightful comments. This research has been supported in part by observations made with the Nordic Optical Telescope, operated by the Nordic Optical Telescope Scientific Association at the Observatorio del Roque de los Muchachos, La Palma, Spain, of the Instituto de Astrofisica de Canarias.  G.P. acknowledges  support from UN (project HERMES 45442). J.H. acknowledges support from the National Research Council of México (CONACyT) project No. 86372 and the PAPIIT UNAM projects IA102921 and IN103320. E.M.M. acknowledges the receipt of the grant from the Abdus Salam International Centre for Theoretical Physics, Trieste, Italy. C.R.Z. acknowledges support from project CONACYT CB2018 A1-S-9754, M\'exico. A.R.L. acknowledges financial support provided in Chile by Comisi\'on Nacional de Investigaci\'on Cient\'ifica y Tecnol\'ogica (CONICYT) through the FONDECYT project 1170476 and by the QUIMAL project 130001. \\

\bibliography{PinzonetalTEX}{}
\bibliographystyle{aasjournal}

\startlongtable
\begin{deluxetable*}{cccccccccc}
\tablecaption{ Properties of the HAeBes sample. Columns 1 and 2 shows the target names, columns 3 and 4 are the RA and DEC, respectively, column 5 is the log\Teff, column 6 the spectral type, column 7 is the visual extinction, column 8 gives the OB1 subgroup, column 9 the distance reported by \cite{briceno2019}  whereas column 10 correspond the distance from Gaia-EDR3 .}  \label{tab:table1}
\tablehead{
\colhead{HIP} & \colhead{HD} & \colhead{RA}  &   \colhead{DEC}      &\colhead{log[\Teff(K)]\tablenotemark{a}} &  \colhead{SpT\tablenotemark{a}}  & \colhead{\Av (mag)\tablenotemark{a}} &\colhead{OB1}   & \colhead{d (pc)\tablenotemark{a}}& \colhead{Gaia  distance (pc)\tablenotemark{b}}  }
\startdata
26752   &   37806    &   05 41 02.29    &   -02 43 0.7     & 4.01 & B9 & 0.21$\pm$0.18   &  bc & 400 & 397$\pm$4 \\ 
25299   &   287841   &   05 24 42.80   &   +01 43 48.2    & 3.88 & A8 &0.0$\pm$0.42  &  a  & 360 & 336$\pm$2 \\
25258   &   287823  &    05 24 08.05   &   +02 27 46.9     & 3.94  &A3 &0.45$\pm$0.23  &  a  & 360 & 343$\pm$3 \\
26500    &  37371   &   05 38 09.90   &   -00 11 1.2         &  3.95 & A2 & 0.23$\pm$0.22 &  bc & 400 & 405$\pm$5 \\
26955   &  38120    &   05 43 11.89   &  -04 59 49.9         &  3.98 & A0&0.13$\pm$0.26 &  bc & 400 & 381$\pm$5 \\
27059   &   38238   &   10 07 19.95     &  +16 45 45.6    	    &  3.87 &A9 & 0.37$\pm$0.27 &  a & 400 & 323$\pm$3 \\ \hline
\enddata
\tablenotetext{a}{\cite{briceno2019}}
\tablenotetext{b}{\cite{gaia2020}}
\end{deluxetable*}

\startlongtable
\begin{deluxetable*}{ccccccccc}
  \tablecaption{ Standard stars properties. Columns 1 and 2 correspond to identifiers, columns 3 and 4 are the RA and DEC., column 5 is the log\Teff, column 6 the spectral type, column 7 is the heliocentric radial velocity, column 8 the projected rotational velocity and column 9 the Gaia-EDR3 distance.}  \label{tab:table2}
\tablehead{
\colhead{HIP} & \colhead{HD} & \colhead{RA}  &   \colhead{DEC}      &\colhead{log[\Teff(K)]} &\colhead{SpT} &\colhead{RV (km $s^{-1}$) \tablenotemark{b}}  & \colhead{$v$sin$i$ (km $s^{-1}$) }& \colhead{Gaia  distance (pc)\tablenotemark{d}}  }
\startdata
49583        &   87737   &   10 07 19.95   &  +16 45 45.5    	    &  3.99 & A0Ib &  +3.3  & 23\tablenotemark{a} & 556$\pm$92 \\ 
  34045      &   53244   &   07 03 45.49   &  -15 37 59.8     	    &  4.13 & B8II & +32  & 36\tablenotemark{c} & 132$\pm$4 \\ \hline
\enddata

\tablenotetext{a}{\cite{royer2002}}
\tablenotetext{b}{\cite{duflot1995}}
\tablenotetext{c}{\cite{simon2017}}
\tablenotetext{d}{\cite{gaia2020}}
\end{deluxetable*}

\begin{longrotatetable}
\begin{deluxetable*}{ccccccccc}
\tablecaption{ Derived stellar parameters for confirmed members in the $\sigma$-Orionis cluster reported by \citep{hernandez2014} and with errors in Gaia-EDR3 parallaxes below 0.2\%. Column 1 corresponds to the 2MASS identifier,  column 2 is the effective temperature, columns 3, 4 and 5 give  mass, radius and age, respectively. The spectral types are shown in column 6.  The $v$sin$i$ obtained through CCF analysis are shown in column 7 whereas those obtained from Fourier are indicated in column 8. Finally the flag (Y/N) in column 9 corresponds to a previous classification based on a $H\alpha$ line.}  \label{tab:tableTTs}
\tablehead{
\colhead{2MASS-ID}   &  \colhead{{\Teff} (K)} & \colhead{$M/M_{\odot}$} &\colhead{Age ($\times 10^6$y) }& \colhead{$R/R_{\odot}$} & 
\colhead{SpT}\tablenotemark{a} & \colhead{$v$sin$i^{CCF}$ (km $s^{-1}$)}    & \colhead{$v$sin$i^{Fourier}$ (km $s^{-1}$)}   &  \colhead{Acc?} }
\startdata
05393511-0247299 &4268$\pm$256 &0.93$^{0.20}_{-0.20}$ &3.09$^{6.92}_{-1.82}$ &1.58$\pm$0.13 &K5.5 &43$\pm$2.7 &43.5$\pm$1.7 &...\\
05380649-0228494 &6665$\pm$279 &1.82$^{0.03}_{-0.04}$ &6.76$^{7.24}_{-6.17}$ &2.72$\pm$0.18 &F3.5 &55$\pm$9 &59.3$\pm$1.4 &...\\
05400696-0228300 &6231$\pm$181 &1.39$^{0.03}_{-0.03}$ &12.88$^{13.49}_{-11.48}$ &1.79$\pm$0.09 &F7.5 &17.1$\pm$4 &15$\pm$6 &...\\
05385911-0247133 &3711$\pm$101 &0.40$^{0.04}_{-0.03}$ &0.13$^{0.19}_{-0.13}$ &3.17$\pm$0.20 &M2.0 &24.2$\pm$3.3 &21.6$\pm$4.5 &...\\
05393654-0242171 &5935$\pm$135 &2.47$^{0.15}_{-0.12}$ &2.51$^{3.02}_{-2.04}$ &4.11$\pm$0.17 &G1.0 &200$\pm$10 &196.7$\pm$3.05 &...\\
05374963-0236182 &5855$\pm$323 &2.13$^{0.17}_{-0.20}$ &3.55$^{5.13}_{-2.40}$ &3.17$\pm$0.26 &G2.5 &$<$9.0 &10.5$\pm$3.5 &...\\
05375440-0239298 &5780$\pm$146 &2.23$^{0.13}_{-0.05}$ &3.09$^{3.31}_{-2.34}$ &3.30$\pm$0.14 &G5.0 &23.3$\pm$0.6 &21.2$\pm$3.2 &...\\
05391717-0225433 &3645$\pm$120 &0.46$^{0.11}_{-0.09}$ &2.63$^{4.37}_{-2.00}$ &1.27$\pm$0.09 &M1.5 &$<$9.0 &14.1$\pm$8.5 &...\\
05372831-0224182 &3477$\pm$120 &0.35$^{0.08}_{-0.06}$ &18.62$^{25.70}_{-14.13}$ &0.62$\pm$0.04 &M3.0 &$<$9.0 &11.1$\pm$3.9 &Y\\
05373666-0234003 &3471$\pm$56 &0.35$^{0.04}_{-0.03}$ &4.37$^{5.01}_{-3.72}$ &1.00$\pm$0.03 &M3.0 &33.8$\pm$5.6 &34.1$\pm$10 &Y\\
05373784-0245442 &3578$\pm$68 &0.41$^{0.05}_{-0.04}$ &2.46$^{3.02}_{-2.14}$ &1.30$\pm$0.05 &M2.0 &23.5$\pm$0.4 &20.1$\pm$4 &N\\
05374527-0228521 &3303$\pm$138 &0.25$^{0.07}_{-0.08}$ &5.50$^{8.71}_{-2.82}$ &0.81$\pm$0.05 &M4.5 &... &16.5$\pm$3.7 &Y\\
05375161-0235257 &3579$\pm$62 &0.39$^{0.04}_{-0.03}$ &1.05$^{1.20}_{-0.98}$ &1.73$\pm$0.06 &M2.0 &34.9$\pm$4.6 &36.5$\pm$1.4 &N\\
05375404-0244407 &3485$\pm$242 &0.35$^{0.17}_{-0.14}$ &3.02$^{5.50}_{-1.55}$ &1.16$\pm$0.12 &M3.0 &23.3$\pm$0.6 &21.2$\pm$3.2 &N\\
05380055-0245097 &3237$\pm$79 &0.21$^{0.04}_{-0.04}$ &1.32$^{1.62}_{-1.05}$ &1.29$\pm$0.05 &M5.0 &47$\pm$6.1 &44$\pm$2.4 &Y\\
05380897-0220109 &3424$\pm$55 &0.32$^{0.03}_{-0.03}$ &3.47$^{4.07}_{-2.88}$ &1.05$\pm$0.04 &M3.5 &... &18.6$\pm$4.3 &Y\\
05381718-0222256 &3233$\pm$80 &0.21$^{0.04}_{-0.04}$ &3.39$^{4.79}_{-2.24}$ &0.88$\pm$0.04 &M5.0 &$<$9.0 &6.9$\pm$3.8 &Y\\
05382354-0241317 &3310$\pm$72 &0.25$^{0.05}_{-0.04}$ &3.16$^{4.17}_{-2.29}$ &0.98$\pm$0.04 &M4.5 &$<$9.0 &12.9$\pm$2.2 &Y\\
05382774-0243009 &3460$\pm$106 &0.33$^{0.07}_{-0.05}$ &1.02$^{1.18}_{-0.85}$ &1.67$\pm$0.09 &M3.0 &29.1$\pm$3.8 &26.7$\pm$1.4 &Y\\
05383284-0235392 &4206$\pm$162 &0.81$^{0.15}_{-0.13}$ &1.48$^{2.19}_{-1.05}$ &1.91$\pm$0.10 &K6.0 &22.8$\pm$4.2 &23.6$\pm$0.4 &N\\
05383546-0231516 &4044$\pm$178 &0.64$^{0.16}_{-0.15}$ &0.85$^{1.32}_{-0.56}$ &2.14$\pm$0.13 &K7.0 &21.9$\pm$2.9 &23.4$\pm$0.4 &N\\
05383745-0250236 &3302$\pm$186 &0.24$^{0.11}_{-0.10}$ &1.59$^{2.51}_{-0.34}$ &1.27$\pm$0.11 &M4.5 &25.9$\pm$7.4 &23.7$\pm$8 &Y\\
05384008-0250370 &3288$\pm$122 &0.24$^{0.08}_{-0.06}$ &3.98$^{6.17}_{-2.46}$ &0.88$\pm$0.06 &M4.5 &$<$9.0 &8.3$\pm$3.7 &Y\\
05384129-0237225 &3855$\pm$168 &0.53$^{0.17}_{-0.08}$ &0.85$^{1.32}_{-0.65}$ &2.04$\pm$0.12 &M0.0 &14.5$\pm$1.5 &17.9$\pm$1.5 &N\\
05384355-0233253 &3674$\pm$207 &0.46$^{0.16}_{-0.11}$ &1.51$^{2.88}_{-1.05}$ &1.56$\pm$0.14 &M1.5 &19$\pm$4 &21.6$\pm$1.7 &N\\
05384993-0241228 &3458$\pm$114 &0.34$^{0.08}_{-0.06}$ &2.24$^{2.75}_{-1.78}$ &1.25$\pm$0.07 &M3.0 &23.1$\pm$4.5 &21.4$\pm$3.5 &N\\
05385317-0243528 &3712$\pm$148 &0.48$^{0.12}_{-0.08}$ &1.41$^{2.09}_{-1.05}$ &1.65$\pm$0.09 &M1.0 &19.5$\pm$2.1 &23.1$\pm$0.5 &N\\
05385831-0216101 &3850$\pm$94 &0.60$^{0.13}_{-0.05}$ &2.69$^{3.63}_{-2.19}$ &1.44$\pm$0.06 &M0.0 &21$\pm$3.1 &23.6$\pm$0.5 &N\\
05390276-0229558 &3371$\pm$123 &0.28$^{0.06}_{-0.08}$ &1.62$^{2.14}_{-1.15}$ &1.31$\pm$0.08 &M4.0 &20.1$\pm$2.5 &19.3$\pm$0.9 &N\\
05390853-0251465 &3776$\pm$74 &0.51$^{0.05}_{-0.04}$ &1.26$^{1.62}_{-1.07}$ &1.74$\pm$0.08 &M0.5 &... &22.7$\pm$0.6 &N\\
05392286-0233330 &3594$\pm$120 &0.42$^{0.10}_{-0.07}$ &2.63$^{3.89}_{-2.09}$ &1.27$\pm$0.07 &M2.0 &$<$9.0 &14.5$\pm$2.9 &N\\
05392456-0220441 &4175$\pm$319 &0.81$^{0.32}_{-0.28}$ &1.86$^{4.47}_{-0.89}$ &1.77$\pm$0.18 &K6.0 &51.9$\pm$5.1 &53.2$\pm$2.8 &N\\
05392650-0252152 &3367$\pm$59 &0.29$^{0.03}_{-0.04}$ &2.57$^{3.09}_{-2.04}$ &1.11$\pm$0.04 &M4.0 &22.4$\pm$3.5 &18$\pm$1 &Y\\
05393291-0247492 &3782$\pm$70 &0.50$^{0.04}_{-0.04}$ &0.93$^{1.10}_{-0.79}$ &1.93$\pm$0.06 &M0.5 &31.6$\pm$2.8 &27.2$\pm$7.8 &N\\
05393729-0226567 &3964$\pm$168 &0.62$^{0.18}_{-0.13}$ &1.18$^{1.91}_{-0.81}$ &1.89$\pm$0.11 &K7.5 &28.6$\pm$4 &29.6$\pm$3.2 &N\\
05382119-0254110 &3148$\pm$90 &0.17$^{0.05}_{-0.03}$ &1.32$^{1.78}_{-0.49}$ &1.21$\pm$0.05 &M5.5 &10.2$\pm$2.3 &17.2$\pm$0.6 &Y\\
05385410-0249297\tablenotemark{b} &5080$\pm$256 &1.51$^{0.03}_{-0.08}$ &4.68$^{6.76}_{-2.40}$ &1.95$\pm$0.15 &K1.0 &31$\pm$2 &29.1$\pm$0.6 &N\\
05391163-0236028\tablenotemark{b} &4077$\pm$167 &0.72$^{0.13}_{-0.19}$ &1.62$^{2.24}_{-0.98}$ &1.77$\pm$0.10 &K7.0 &... &... &N\\
05393256-0239440\tablenotemark{b}&4053$\pm$170 &0.60$^{0.15}_{-0.12}$ &0.42$^{0.59}_{-0.32}$ &2.74$\pm$0.16 &K7.0 &29.9$\pm$7.1 &35.2$\pm$3.1 &N\\
05384027-0230185\tablenotemark{b} &4057$\pm$159 &0.64$^{0.14}_{-0.12}$ &0.68$^{0.96}_{-0.49}$ &2.31$\pm$0.12 &K7.0 &21$\pm$1.6 &17.7$\pm$1.7 &N\\
05375486-0241092 &3239$\pm$84 &0.21$^{0.04}_{-0.05}$ &3.09$^{4.27}_{-2.09}$ &0.91$\pm$0.04 &M5.0 &$<$9.0 &8.2$\pm$2.7 &Y\\
05381886-0251388 &3530$\pm$119 &0.38$^{0.07}_{-0.06}$ &2.19$^{2.88}_{-1.74}$ &1.31$\pm$0.08 &M2.5 &$<$9.0 &13.2$\pm$2.4 &N\\
05384423-0240197 &4208$\pm$71 &0.81$^{0.07}_{-0.08}$ &1.41$^{1.70}_{-1.15}$ &1.95$\pm$0.05 &K6.0 &22.6$\pm$3.2 &23.4$\pm$0.6 &N\\
05382911-0236026 &3659$\pm$63 &0.46$^{0.04}_{-0.05}$ &2.19$^{2.63}_{-1.82}$ &1.39$\pm$0.04 &M1.5 &... &23.3$\pm$0.6 &N\\
05383431-0235000 &4054$\pm$170 &0.65$^{0.14}_{-0.15}$ &0.81$^{1.20}_{-0.55}$ &2.16$\pm$0.12 &K7.0 &32.4$\pm$3.1 &31.9$\pm$0.8 &N\\
05373094-0223427 &3585$\pm$61 &0.42$^{0.06}_{-0.04}$ &3.89$^{4.90}_{-3.47}$ &1.11$\pm$0.04 &M2.0 &18.4$\pm$6.5 &21.1$\pm$2 &Y\\
05380107-0245379\tablenotemark{b} &3307$\pm$66 &0.25$^{0.04}_{-0.03}$ &1.91$^{2.46}_{-1.55}$ &1.16$\pm$0.06 &M4.5 &27.9$\pm$7.5 &30.3$\pm$1.4 &Y\\
05380674-0230227 &3867$\pm$89 &0.55$^{0.08}_{-0.06}$ &1.07$^{1.35}_{-0.87}$ &1.89$\pm$0.07 &M0.0 &16.6$\pm$2.1 &20.3$\pm$1.2 &Y\\
05380994-0251377 &3473$\pm$50 &0.31$^{0.02}_{-0.02}$ &0.13$^{0.13}_{-0.13}$ &2.78$\pm$9.56 &M3.0 &$<$9.0 &9.9$\pm$3.5 &Y\\
05381315-0245509 &3660$\pm$66 &0.43$^{0.04}_{-0.04}$ &1.05$^{1.20}_{-0.89}$ &1.77$\pm$0.05 &M1.5 &23.1$\pm$0.5 &21.1$\pm$3.1 &Y\\
05381319-0226088 &3361$\pm$124 &0.27$^{0.07}_{-0.07}$ &1.18$^{1.74}_{-0.81}$ &1.46$\pm$0.13 &M4.0 &22.1$\pm$0.7 &19.2$\pm$5.5 &Y\\
05382050-0234089 &3369$\pm$124 &0.27$^{0.05}_{-0.08}$ &0.74$^{0.89}_{-0.21}$ &1.83$\pm$0.12 &M4.0 &24.4$\pm$6.3 &23.4$\pm$0.5 &Y\\
05382358-0220475 &3362$\pm$129 &0.29$^{0.06}_{-0.10}$ &8.71$^{16.98}_{-3.89}$ &0.71$\pm$0.11 &M4.0 &$<$9.0 &11.9$\pm$3.7 &Y\\
05382543-0242412 &3290$\pm$222 &0.24$^{0.11}_{-0.14}$ &16.22$^{34.67}_{-3.89}$ &0.54$\pm$0.07 &M4.5 &$<$9.0 &13.2$\pm$2.3 &Y\\
05382725-0245096 &4210$\pm$167 &0.93$^{0.09}_{-0.15}$ &5.01$^{7.59}_{-3.24}$ &1.37$\pm$0.07 &K6.0 &19.2$\pm$2.7 &21.9$\pm$6.8 &Y\\
05383368-0244141 &4471$\pm$261 &0.98$^{0.40}_{-0.28}$ &0.54$^{0.98}_{-0.29}$ &2.97$\pm$0.33 &K4.5 &24.2$\pm$7.3 &22.7$\pm$0.6 &Y\\
05384301-0236145 &3716$\pm$142 &0.47$^{0.12}_{-0.07}$ &1.15$^{1.74}_{-0.89}$ &1.75$\pm$0.10 &M1.0 &19$\pm$4 &22.9$\pm$0.4 &Y\\
05384537-0241594 &3661$\pm$212 &0.48$^{0.19}_{-0.14}$ &3.89$^{7.59}_{-2.34}$ &1.16$\pm$0.12 &M1.5 &... &22.5$\pm$0.9 &Y\\
05384718-0234368 &3401$\pm$191 &0.30$^{0.09}_{-0.12}$ &1.35$^{1.86}_{-0.44}$ &1.43$\pm$0.12 &M4.0 &$<$9.0 &8.1$\pm$3.7 &Y\\
05390136-0218274 &3780$\pm$77 &0.49$^{0.04}_{-0.04}$ &0.87$^{1.02}_{-0.74}$ &1.96$\pm$0.06 &M0.5 &20.4$\pm$3.2 &19.7$\pm$1 &Y\\
05390297-0241272 &3520$\pm$53 &0.37$^{0.03}_{-0.03}$ &1.29$^{1.41}_{-1.15}$ &1.58$\pm$0.05 &M2.5 &22.9$\pm$3 &18.7$\pm$1.4 &Y\\
05390357-0246269 &3426$\pm$121 &0.32$^{0.06}_{-0.08}$ &2.51$^{3.09}_{-1.66}$ &1.18$\pm$0.07 &M3.5 &17$\pm$4 &20.4$\pm$1.3 &Y\\
05390878-0231115 &3485$\pm$116 &0.35$^{0.06}_{-0.07}$ &2.29$^{2.88}_{-1.74}$ &1.26$\pm$0.07 &M3.0 &14.5$\pm$3.1 &16.9$\pm$2.7 &Y\\
05393982-0231217 &4265$\pm$177 &0.97$^{0.10}_{-0.15}$ &4.57$^{7.59}_{-3.02}$ &1.44$\pm$0.08 &K5.5 &$<$9.0 &15.1$\pm$1.8 &Y\\
05394017-0220480 &4363$\pm$189 &1.00$^{0.18}_{-0.18}$ &2.88$^{5.01}_{-1.74}$ &1.68$\pm$0.10 &K5.0 &21.7$\pm$5.1 &17.8$\pm$0.6 &Y\\
05400889-0233336 &3948$\pm$155 &0.59$^{0.14}_{-0.11}$ &0.87$^{1.29}_{-0.66}$ &2.05$\pm$0.11 &K7.5 &22.04$\pm$4.1 &17.6$\pm$0.9 &Y\\
05383460-0241087 &3575$\pm$123 &0.41$^{0.11}_{-0.07}$ &2.00$^{3.02}_{-1.70}$ &1.37$\pm$0.08 &M2.0 &$<$9.0 &9.6$\pm$2.2 &Y\\
05380826-0235562 &3524$\pm$59 &0.36$^{0.03}_{-0.04}$ &0.98$^{1.07}_{-0.89}$ &1.75$\pm$0.06 &M2.5 &22.7$\pm$4.5 &22.8$\pm$0.5 &Y\\
05381412-0215597\tablenotemark{c} &6260$\pm$226 &1.72$^{0.06}_{-0.02}$ &7.08$^{7.76}_{-6.17}$ &2.50$\pm$0.16 &F7.5 &27.4$\pm$9 &29.1$\pm$7.7 &Y\\
05382684-0238460 &3416$\pm$122 &0.32$^{0.06}_{-0.08}$ &7.59$^{10.47}_{-4.79}$ &0.79$\pm$0.06 &M3.5 &19$\pm$3 &17.4$\pm$1.3 &Y\\
05383587-0243512 &4729$\pm$466 &1.38$^{0.60}_{-0.66}$ &0.78$^{2.82}_{-0.26}$ &3.09$\pm$0.44 &K3.0 &30.5$\pm$1.2 &26.4$\pm$7.3 &Y\\
05375303-0233344\tablenotemark{b} &5899$\pm$124 &1.62$^{0.05}_{-0.05}$ &7.59$^{8.13}_{-6.46}$ &2.19$\pm$0.07 &G2.5 &41.3$\pm$2.5 &46.4$\pm$0.9 &Y\\
05383587-0230433\tablenotemark{b} &4067$\pm$181 &0.68$^{0.16}_{-0.16}$ &1.20$^{1.74}_{-0.78}$ &1.93$\pm$0.12 &K7.0 &84.9$\pm$9.3 &89.3$\pm$1.6 &Y\\
\enddata 
\tablenotetext{a}{\citep{hernandez2014}.}
\tablenotetext{b}{binary candidate identified  by \citet{hernandez2014}.}
\tablenotetext{c}{binary candidate identified  by \citet{kounkel2019}.}

\end{deluxetable*}
\end{longrotatetable}

\startlongtable
\begin{deluxetable*}{ccccccccc}
\tablecaption{ Derived stellar parameters for the HAeBes sample. Column 1 is the identifier, column 2 the radial velocity (RV) obtained from our CCF analysis. RV reported by \citep{alecian2013}  are shown in column 3.  The stellar rotation computed through CCF and Fourier is shown in columns 4 and 5 whereas columns 6, 7 and 8 correspond to the stellar radius, mass and age. The last column gives the WISE infrared intrinsic color [3.4 - 12]$\mu$m. } \label{tab:tablehaebes}
\tablehead{
\colhead{HIP}    &     \colhead{RV (km $s^{-1}$)} & \colhead{RV (km $s^{-1}$)\tablenotemark{a}}  &   \colhead{$v$sin$i^{CCF}$ (km $s^{-1}$)}  & \colhead{$v$sin$i^{Fourier}$  (km $s^{-1}$)\tablenotemark{b}}  &      \colhead{$R/R_{\odot}$}   &  \colhead{$M/M_{\odot}$}   &   \colhead{Age (Myr)}      &\colhead{[3.4-12]$_0$}  }
\startdata
26752     &    17$\pm$1       &  47$\pm$21 &       128$\pm$7 & 120.7$\pm$25.2  &   5.5$^{+0.7}_{-0.5}$   &   4.3$^{+0.2}_{-0.3}$   &   0.85$^{+0.1}_{-0.1}$       & 3.19 \\ 
25299     &   19$\pm$3        &20.0$\pm$3.6      &  118$\pm$18  & 123.5$\pm$9.0      &     1.7$^{+0.2}_{-0.1}$   &   1.7$^{+0.1}_{-0.1}$   &   12.02$^{+1.80}_{-11.47}$    &    2.13 \\
25258A      &   -7.8$\pm$0.3  & -0.3$\pm$1.1    &      10.3$\pm$2.2  & ...      &   2.0$^{+0.0}_{-0.1}$     &  2.1$^{+0.2}_{-0.1}$  &   6.46$^{+0.6}_{-0.5}$         &  2.41 \\
25258B      &   50.5$\pm$2.4   & 54.0$\pm$1.6  &      7.2$\pm$0.3  & ...     &  2.0$^{+0.9}_{-0.1}$    &  2.1$^{+0.6}_{-0.2}$  &   6.46$^{+0.6}_{-0.5}$            &  2.41 \\
26500       &    51$\pm$15   & ...    &116$\pm$15   & 99.4$\pm$5.7        &    3.0$^{+0.1}_{-0.1}$     &   2.5$^{+0.1}_{-0.1}$ &   3.55$^{+0.2}_{-0.2}$            &   1.37   \\
26955     &   26.8$\pm$4.7    &  28$\pm$12 & 103$\pm$5    & 97$\pm$21           &   3.0$^{+0.0}_{-0.1}$    &   2.6$^{+0.1}_{-0.6}$   &   3.31$^{+0.8}_{-0.3}$         & 4.95 \\
27059     &  16.9$\pm$1.5    &  15.0$\pm$2.9            &119$\pm$11    & 111.1$\pm$16.8   &    3.9$^{+0.2}_{-0.1}$   &   2.5$^{+0.1}_{-0.1}$   &  3.02$^{+0.2}_{-0.2}$           &  1.91 \\
\enddata
\tablenotetext{a}{\citep{alecian2013}}
\end{deluxetable*}

\begin{center}
\begin{table*}[h!]
\begin{center}
\begin{tabular}{|l|rr|rr|}
\hline
  & \multicolumn{2}{c|}{\textbf{TTs}} & \multicolumn{2}{c|}{\textbf{HAeBes}} \\
$B_*$ (kG)
& \multicolumn{1}{c} {KS} & \multicolumn{1}{c|} {p-val} & \multicolumn{1}{c} {KS} & \multicolumn{1}{c|}{p-val}\\
\hline
$0.5$     &    $0.60$ & $...$ & $0.85$ & $0.001$\\ 
$1.0$     &    $0.54$ &  $0.002$ & $0.47$ & $0.240$\\
$2.0$     &    $0.20$  & $0.712$  & $0.33$ & $0.631$ \\
$3.0$     &    $0.35$ &   $0.102$ & $0.74$ & $0.008$\\
\hline
\end{tabular}
\caption{\label{tab:ks} KS statistic on observed and expected samples of specific angular momentum as a function on mass.  Expected $J$ were obtained using $\tau_D=3$ Myr. If KS is small or the p-val is high, then we cannot reject the hipothesis that the distributions of the two samples are the same. }
\end{center}
\end{table*}
\end{center}

\begin{figure*}
\gridline{\includegraphics[width=9.0cm]{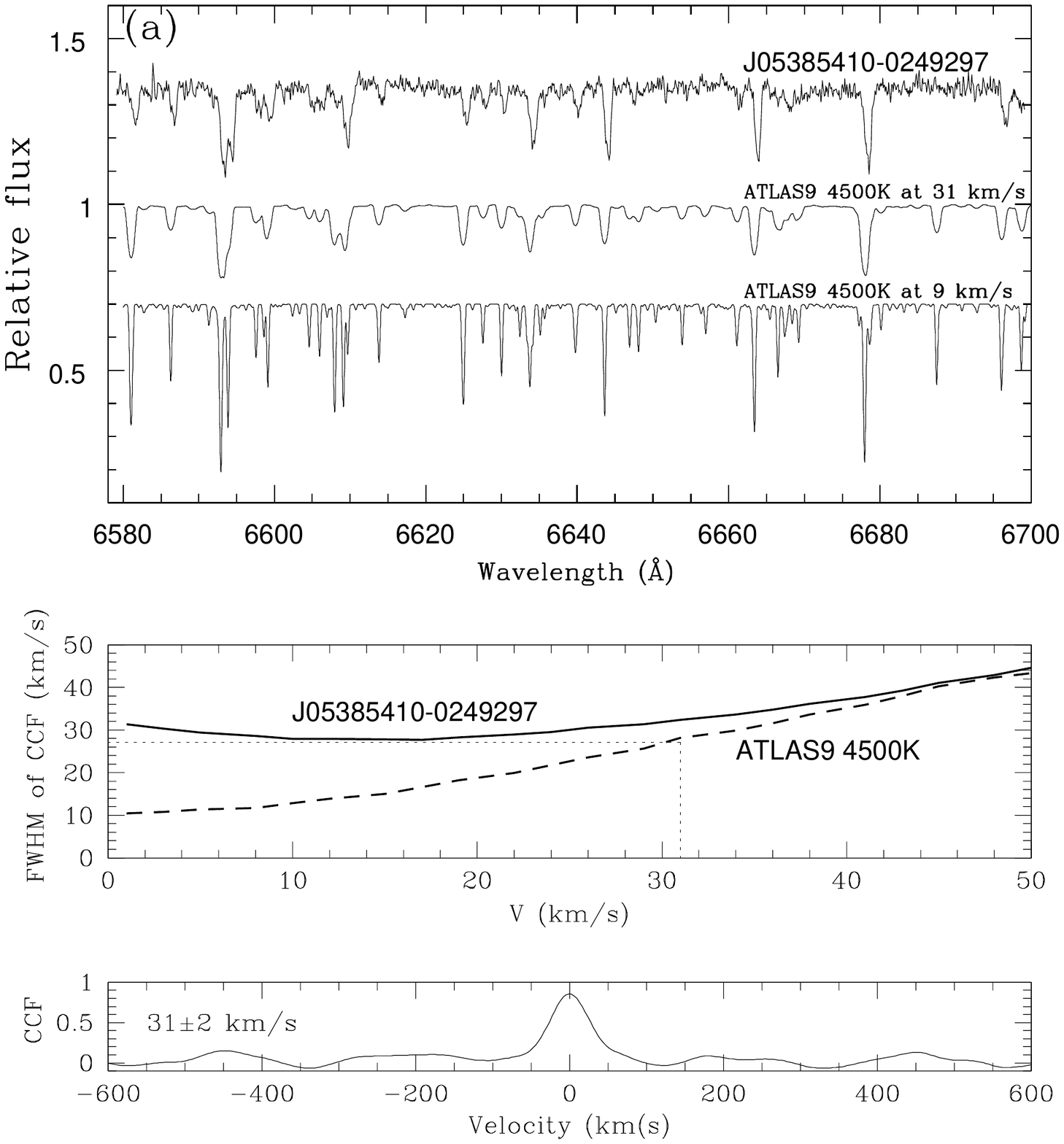}
          \includegraphics[width=9.0cm]{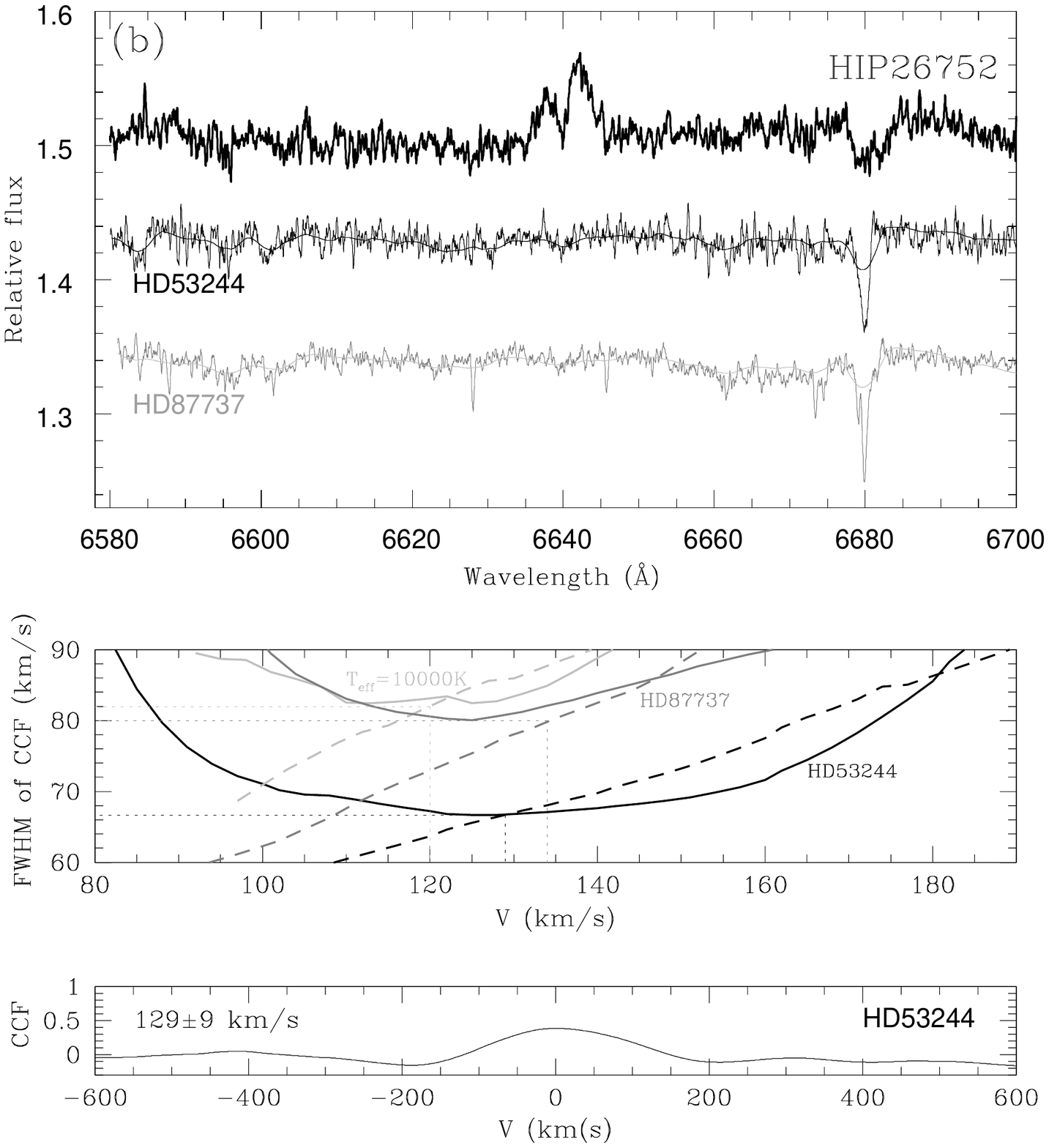}
          }
          \caption{(a) The cross correlation method applied on the TTs 2MASS J05385410-0249297 (SpT K1.0, upper spectrum). The synthetic stellar template (lower spectrum) is shown with its corresponding broadened spectrum in the middle.  In the second panel, the FWHM between object and broadened template is indicated with a solid line whereas dashed line corresponds to the calibration function after interpolation with a polynomial of third order.  The third panel illustrates the CCF corresponding to a broadening of $31$ km $s^{-1}$ which is the velocity that minimizes the width of the CCF using the calibration function.  (b)  Same technique applied to the HAeBe star HIP26752 (SpT B9, upper spectrum) using the spectra of HD53244 (SpT B8II, middle spectrum) and HD87737 (SpT A0Ib, lower spectrum in gray), observed during the same night, as stellar templates. Thin-lines show the broadened spectra at the $v$sin$i$ of the star. The middle panel show the FWHM of the CCF between object and template with thick lines whereas dashed lines represents those computed with the templates HD53244 (black), HD87737 (dark-gray) and synthetic A0 (light-gray). Error bars in the final $v$sin$i$ determination are computed using the $R$ parameter, details are given in the text.}
          \label{fig:f1}
\end{figure*}

\begin{center}
\begin{figure*}
\begin{center}
\includegraphics[width=7.0cm, height=7.0cm]{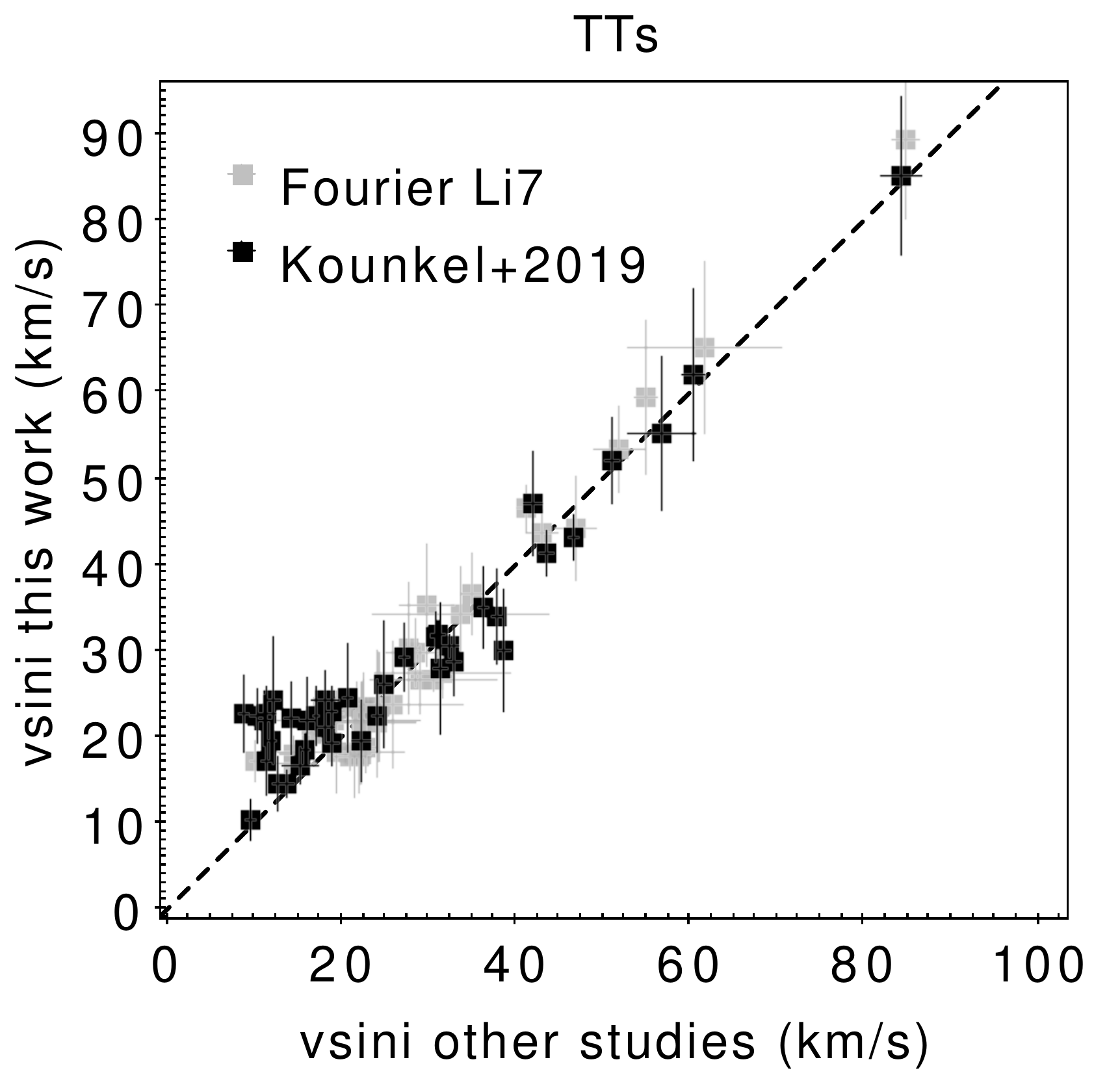}  \includegraphics[width=7.0cm, height=7.0cm]{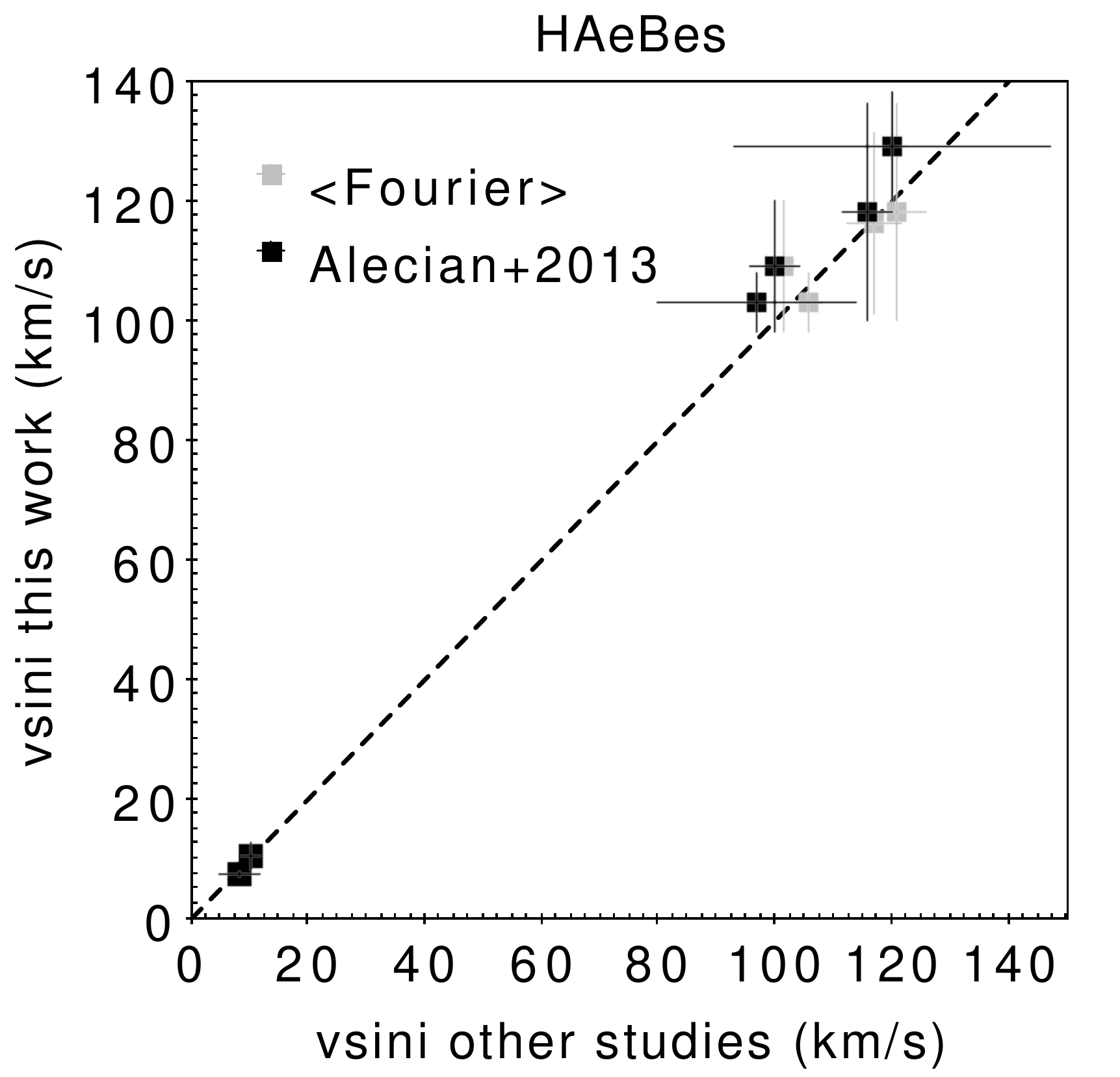}            
\caption{ \textbf{Left.} Comparison  of $v$sin$i$ measurements   using the CCF method with those obtained through FT applied to the Li76707{\AA} absorption line (gray). Symbols in black correspond to a comparison with \citep{kounkel2019}. \textbf{Right.}  Same comparison for HAeBes. Symbols in gray 
 correspond to an average obtained from FT method applied to  MgII4481{\AA}, FeI4226{\AA} and FeI4489{\AA} lines. Symbols in black color  represent a comparison with $v$sin$i$ reported by \citep{alecian2013}.}\label{fig:f2}
\end{center}
\end{figure*}
\end{center}

\begin{figure}
\plotone{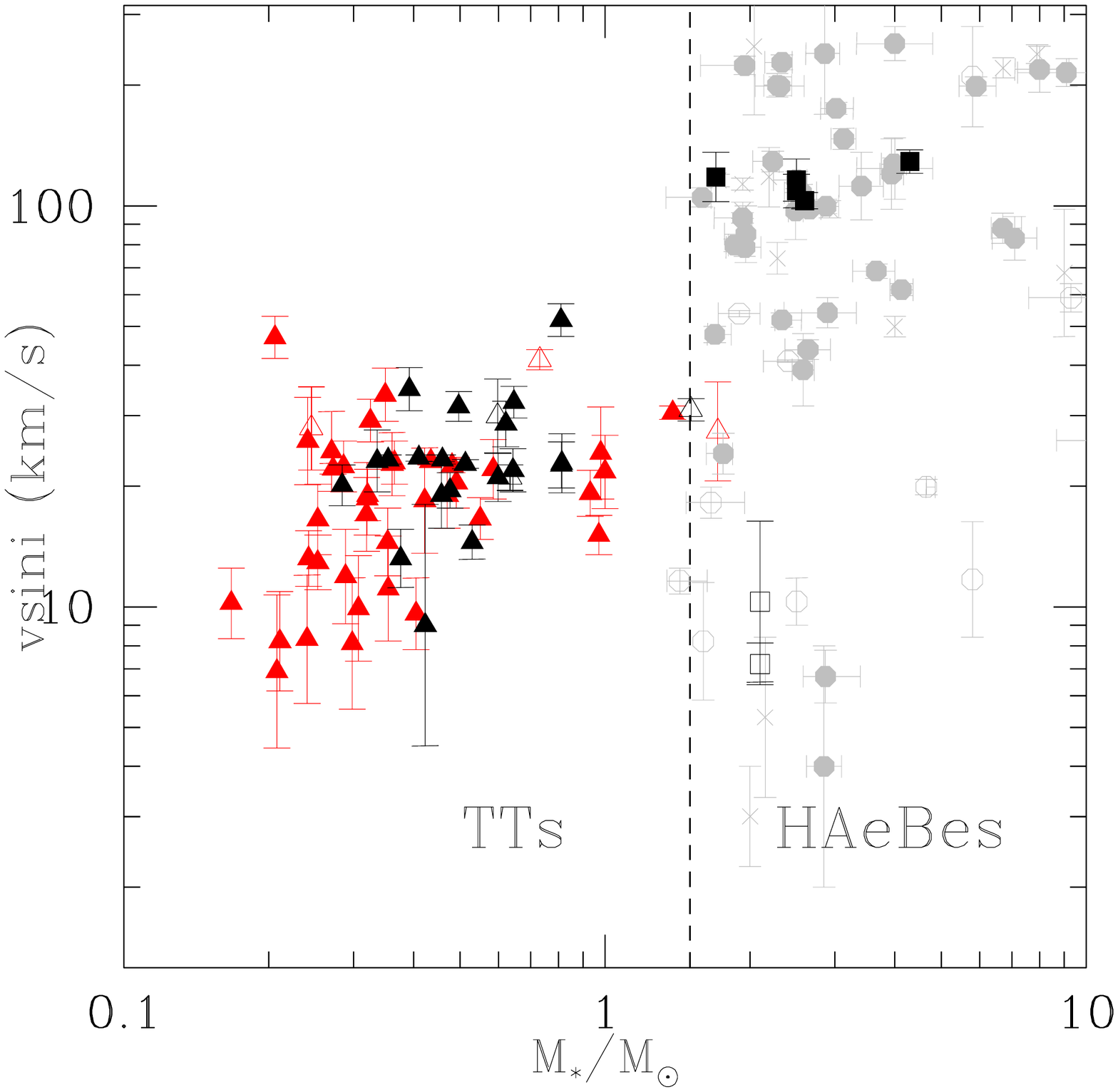}
\caption{ $v$sin$i$ versus stellar mass for TTs (triangles) and HAeBes (rectangles). TTs with active accretion reported by \cite{hernandez2014} are indicated with bigger triangles in red.   Complementary  data of HAeBes taken from \citep{alecian2013} and \citep{fairlamb2015} have been included in gray.  Binaries are indicated with open circles and upper limit values with crosses.   }  
\label{fig:all}
\end{figure}

\begin{figure}
\plotone{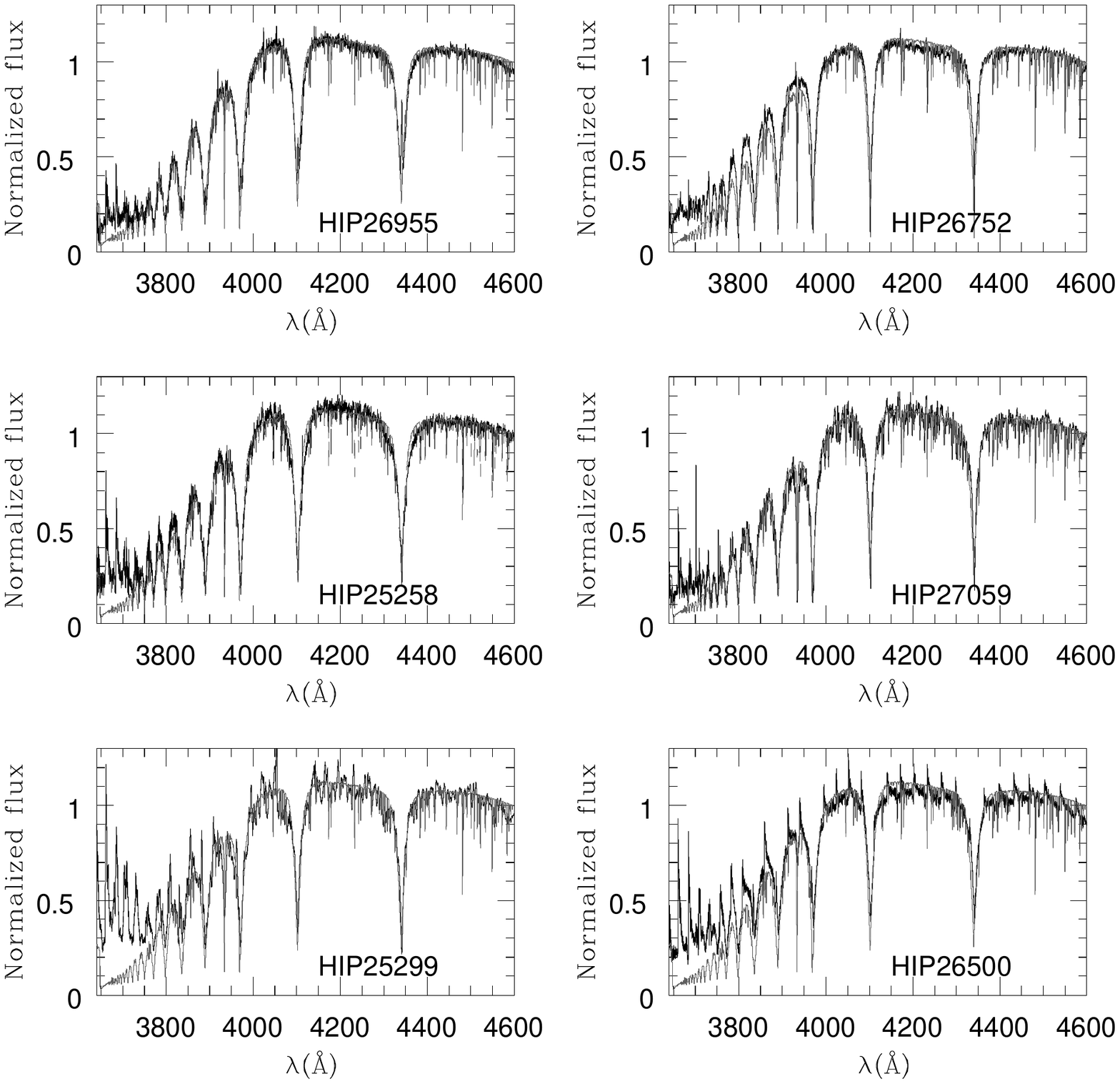}
\caption{Balmer discontinuity in HAeBes. At each panel 
 the star is compared with a synthetic template (gray line) of a A0 star from \cite{kurucz1979}. In spite of contamination from line emission, it is evident the Balmer excess in all objects.  \label{fig:f9}}
\end{figure}

\begin{figure}
\plotone{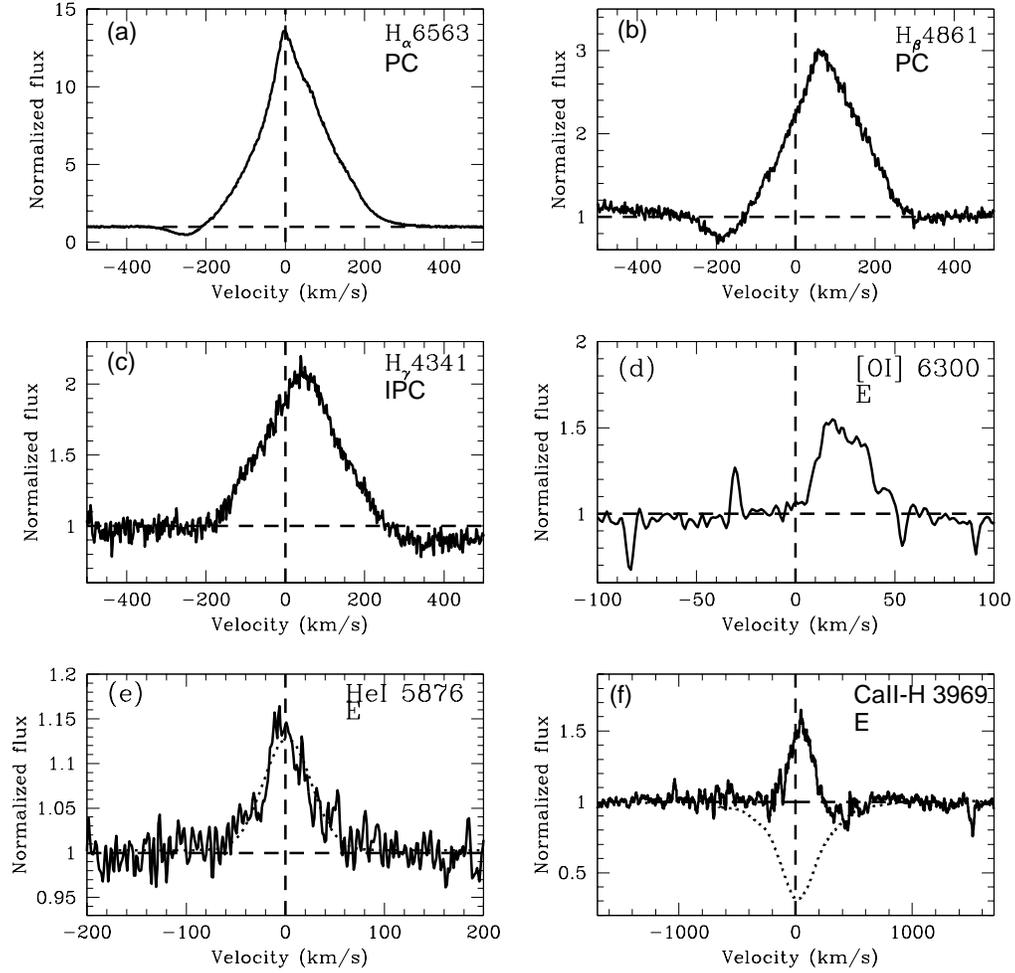}
\caption{Non-photospheric line profiles for HIP26955. The continuum, normalized to 1.0 is marked with a horizontal dashed line while the stellar rest velocity is labeled with a vertical dashed line. Profile classification  and  spectral line are shown at the upper-right part of each panel. (a), (b) and (c) correspond to  $H{\alpha}$ 6563{\AA}, $H{\beta}$ 4861{\AA} and $H{\gamma}$ 4341{\AA}, respectively. (d) The  [OI] 6300.31{\AA}, (e) the HeI 5876{\AA} line and a Gaussian fit (dotted).  (e) the CaII-H 3969{\AA} line and the photospheric contribution (dotted line). The intensity scales were adjusted so that the continuum for all stars would be the same.  \label{fig:buena}}
\end{figure}

\begin{figure}
\plotone{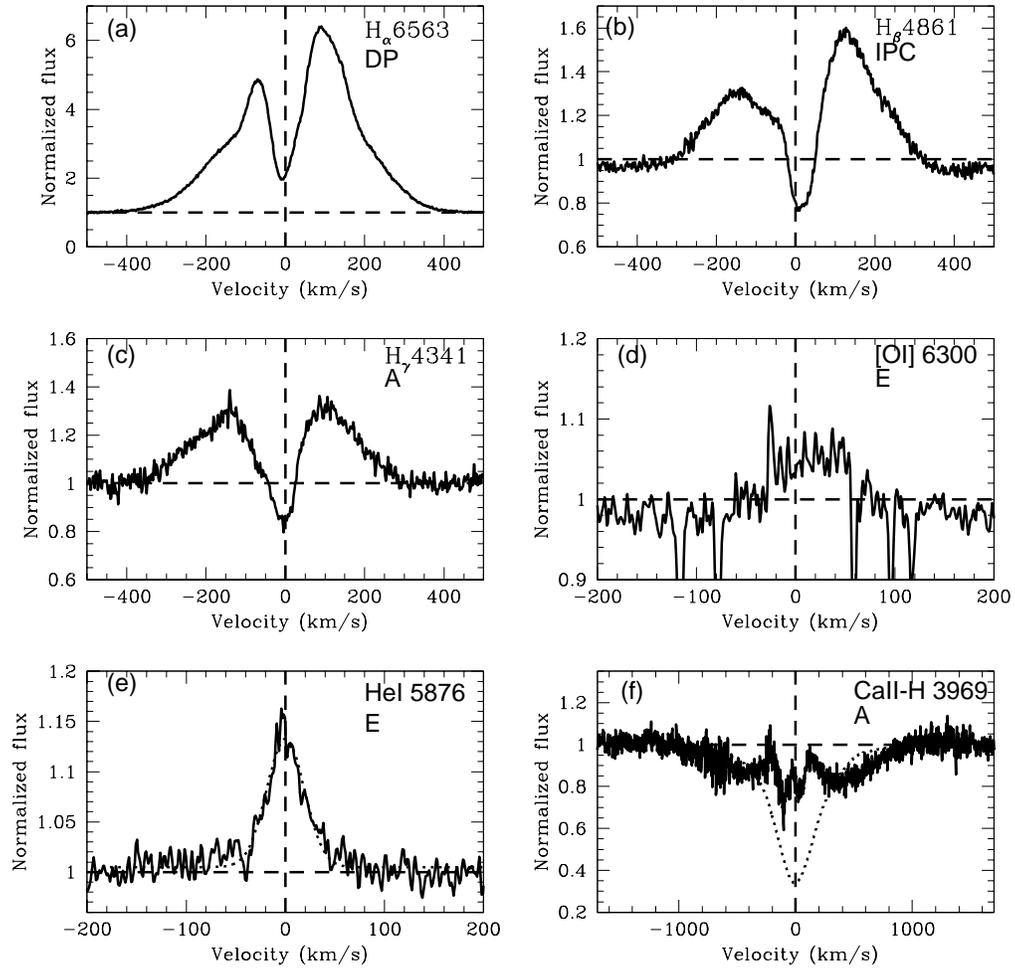}
\caption{Non-photospheric line profiles for HIP26752.  Symbols are the same than Figure \ref{fig:buena}. 
 \label{fig:hip26752}}
\end{figure}

\begin{figure}
\plotone{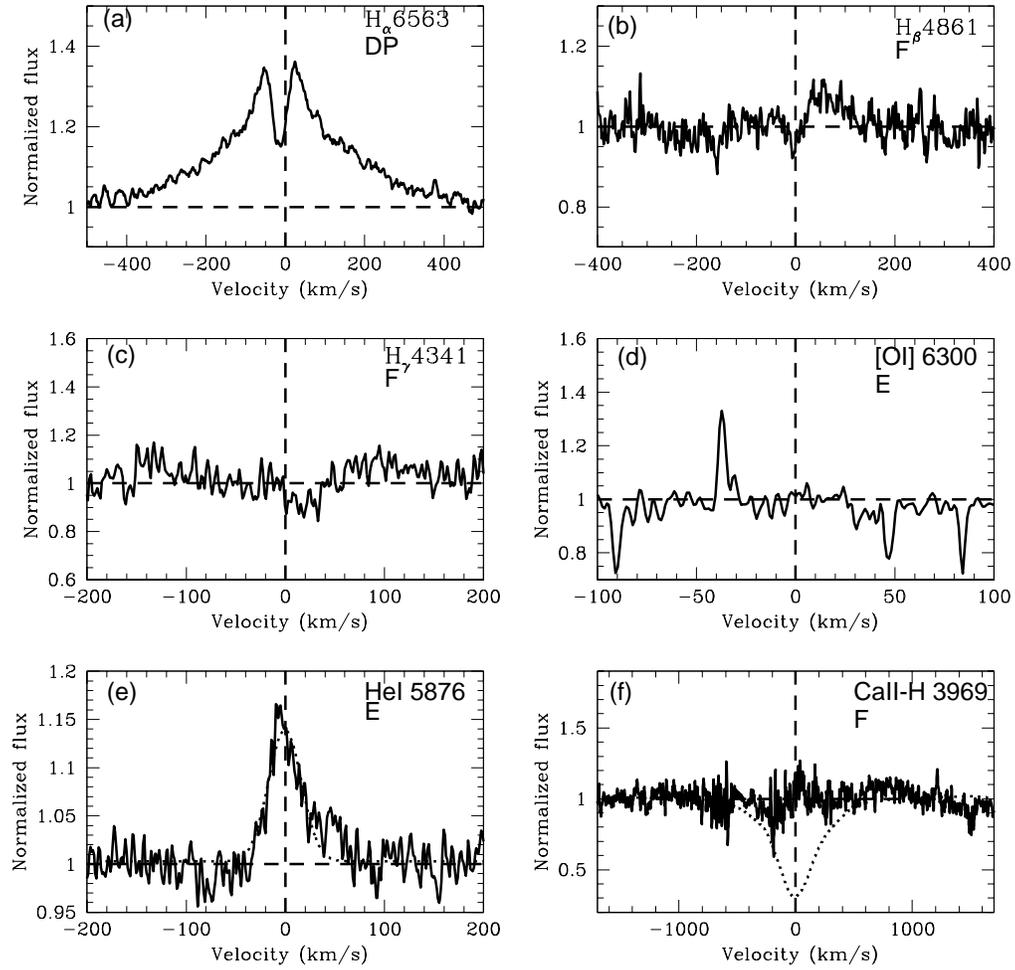}
\caption{same than Figure \ref{fig:buena} but for HIP25258. \label{fig:f4}}
\end{figure}

\begin{figure}
\plotone{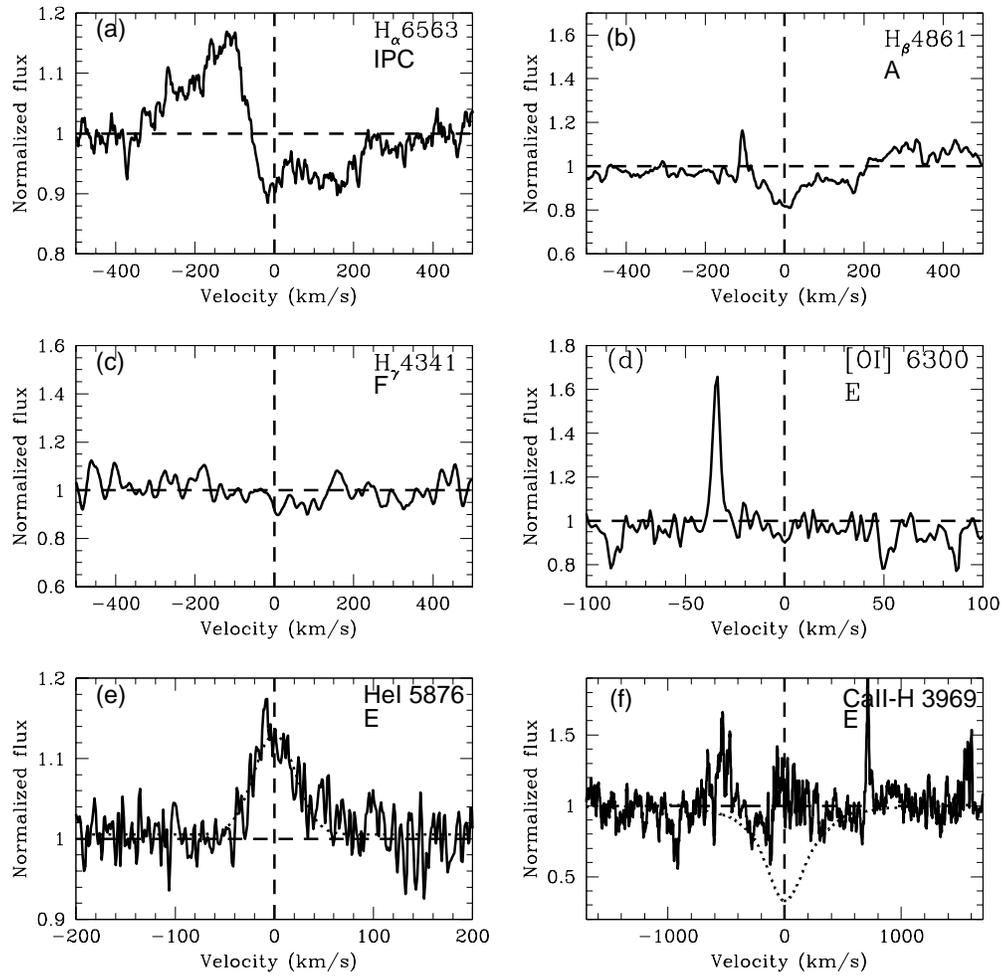}
\caption{same than Figure \ref{fig:buena} but for HIP25299.  \label{fig:f5}}
\end{figure}

\begin{figure}
\plotone{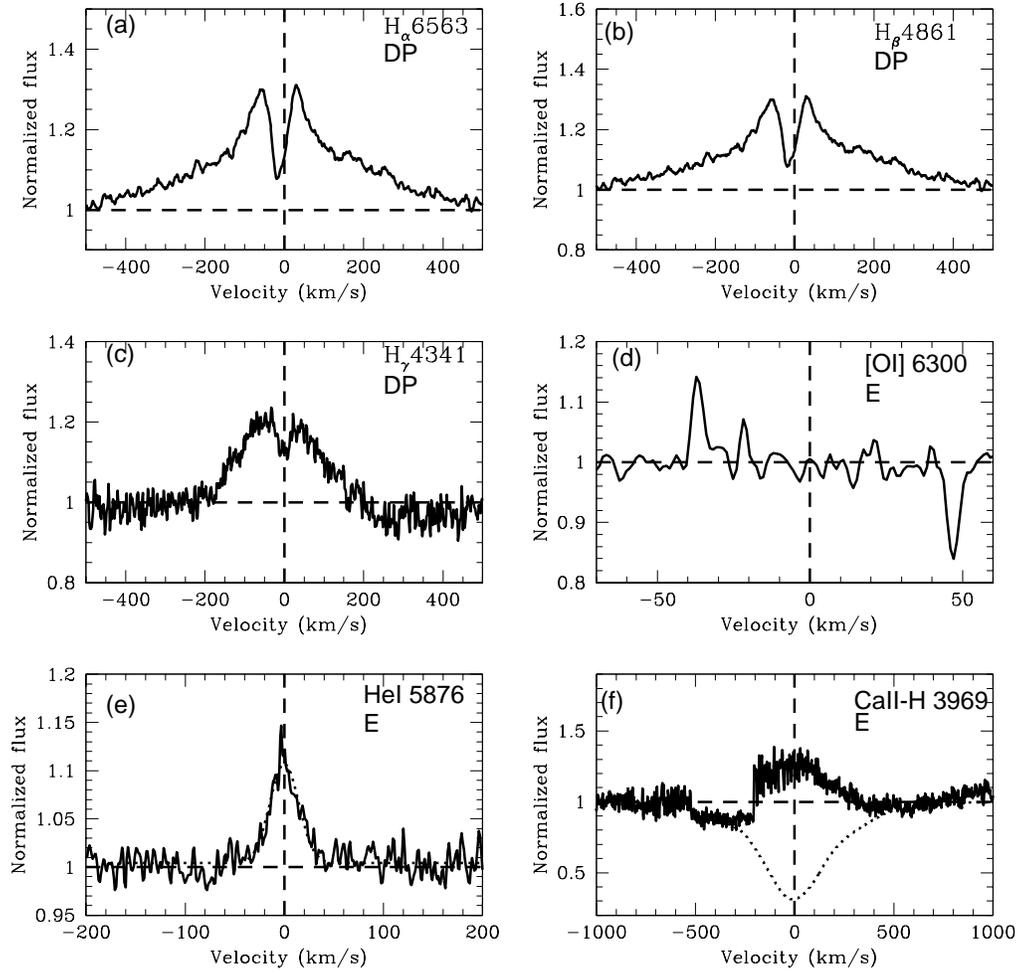}
\caption{same than Figure \ref{fig:buena} but for  HIP26500.  \label{fig:f6}}
\end{figure}

\begin{figure}
\plotone{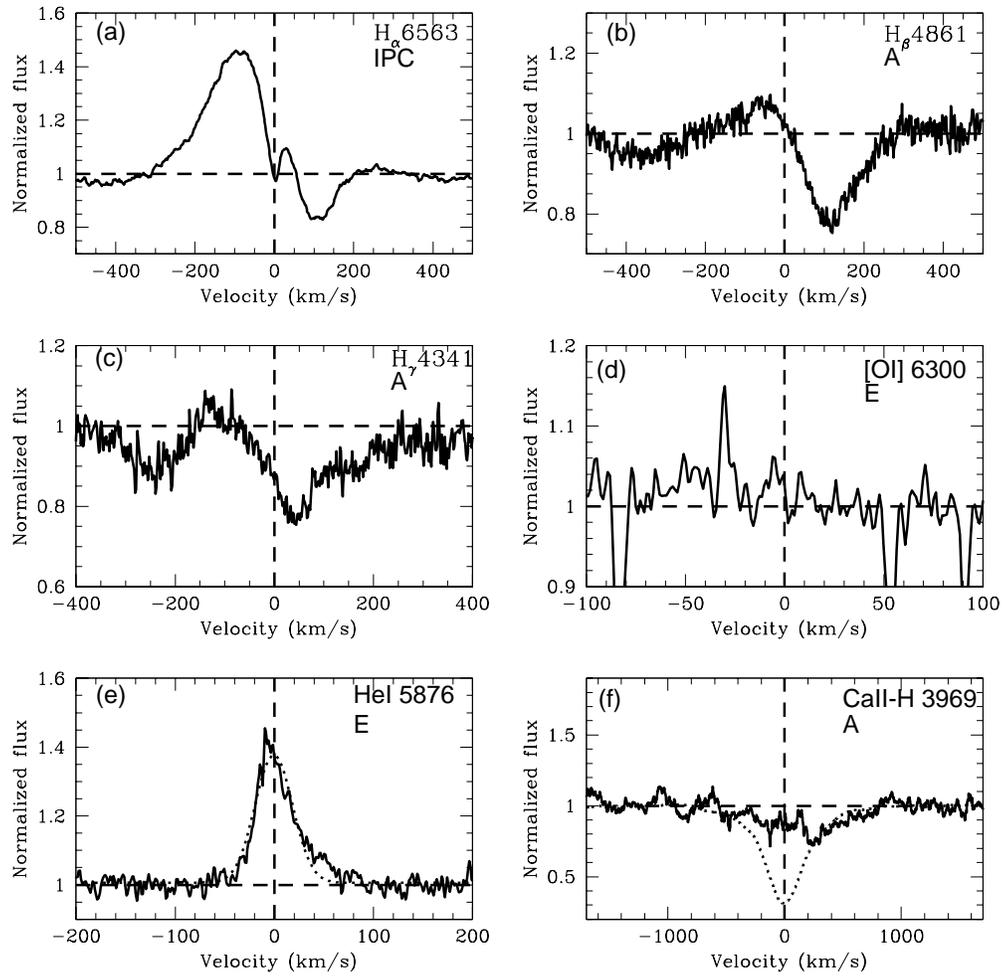}
\caption{same than Figure \ref{fig:buena} but for HIP27059.  \label{fig:f8}}
\end{figure}

\begin{figure}
\plotone{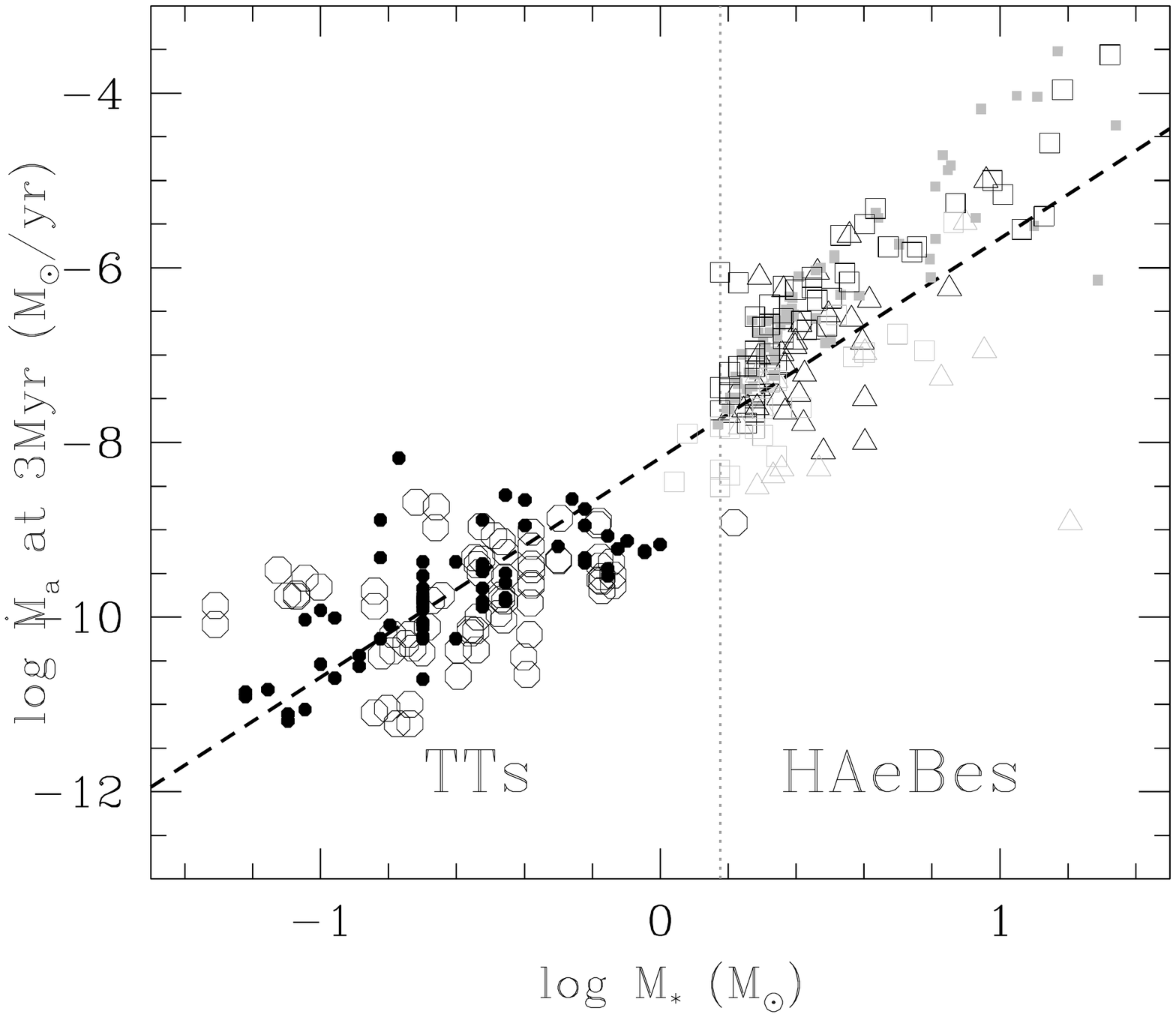}
\caption{ Compilation of stellar accretion at $3$Myr: TTs from \citep{rigliaco2012} (filled circles) and \citep{mauco2016} (open circles).  HAeBes from \citep{fairlamb2015} (open squares in black), \citep{alecian2013} (open triangles in black).  Vertical line represents the limit between TTs and HAeBes whereas the straight line corresponds to the best fit with $\alpha=(2.51\pm0.2)$. This exponent is in agreement with that obtained with the data reported in the recent study by  \citep{wichittanakom2020} (filled squares in gray). We used this  match in order to obtain the  accretion rates at the birth-line (see in the text).}
\label{fig:f10a}
\end{figure}

\begin{figure*}
\plotone{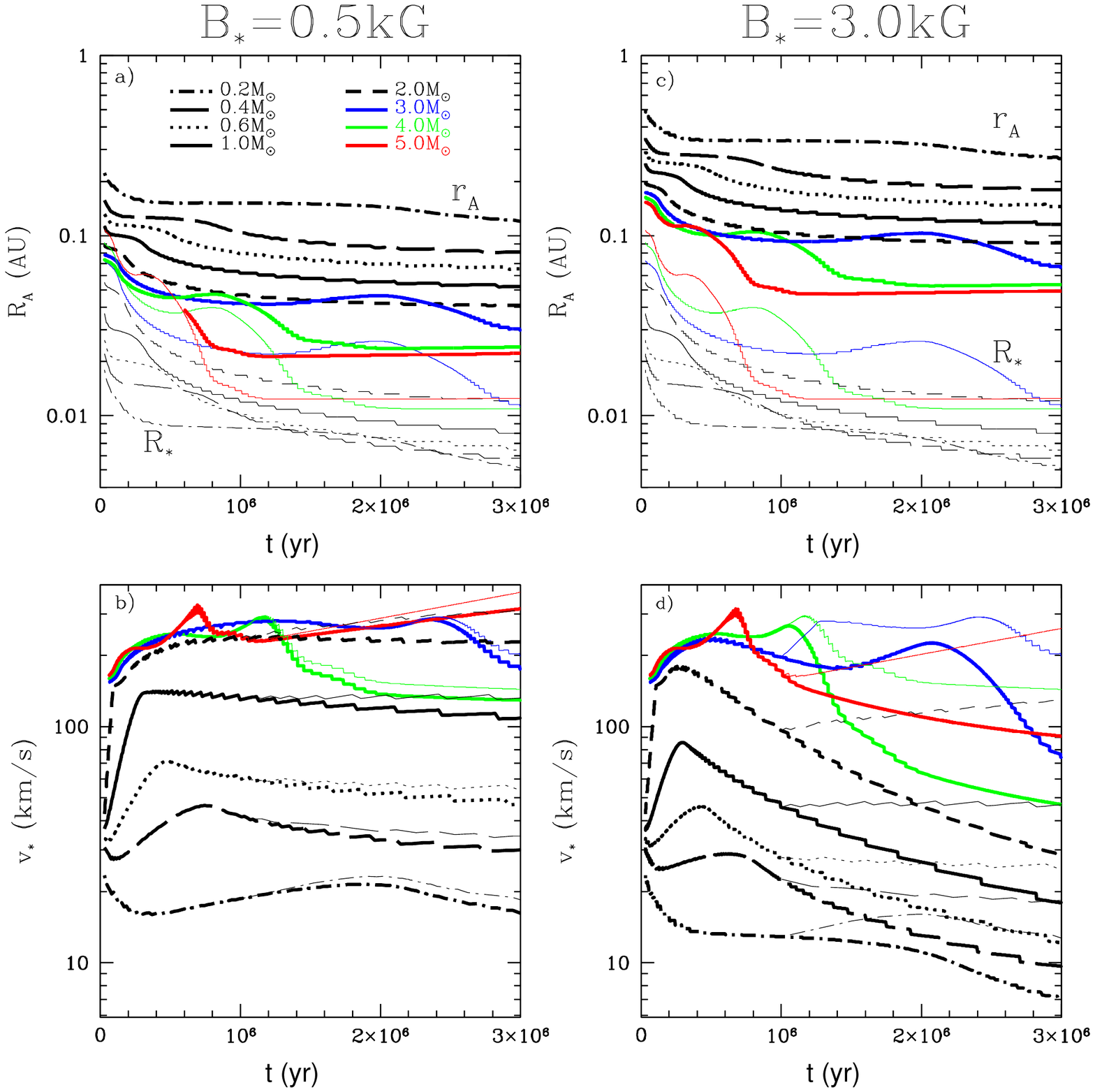}
\caption{ Time evolution of $r_A$ and $v_*$  for two magnetic field strengths.   Panels (a) and (b) show the Alfv\'en radii in astronomical units and stellar rotational velocities in km $s^{-1}$, respectively for 
$B_=0.5$kG. Panels (c) and (d) were obtained using $B_*=3.0$ kG. Thicked lines of a given color or type represent  eight rotational models of protostars with long-lived discs ($\tau_D=3$ Myr) and with masses between $0.2$ and $5.0M_{\odot}$.  Thin lines in panels (b) and (d) represent the rotational evolution for $\tau_D=1$ Myr whereas in panels (a) and (b) thin lines follow the stellar radius evolution as given by \cite{siess2000}.
\label{fig:f12}}
\end{figure*}

\begin{figure*}
\plotone{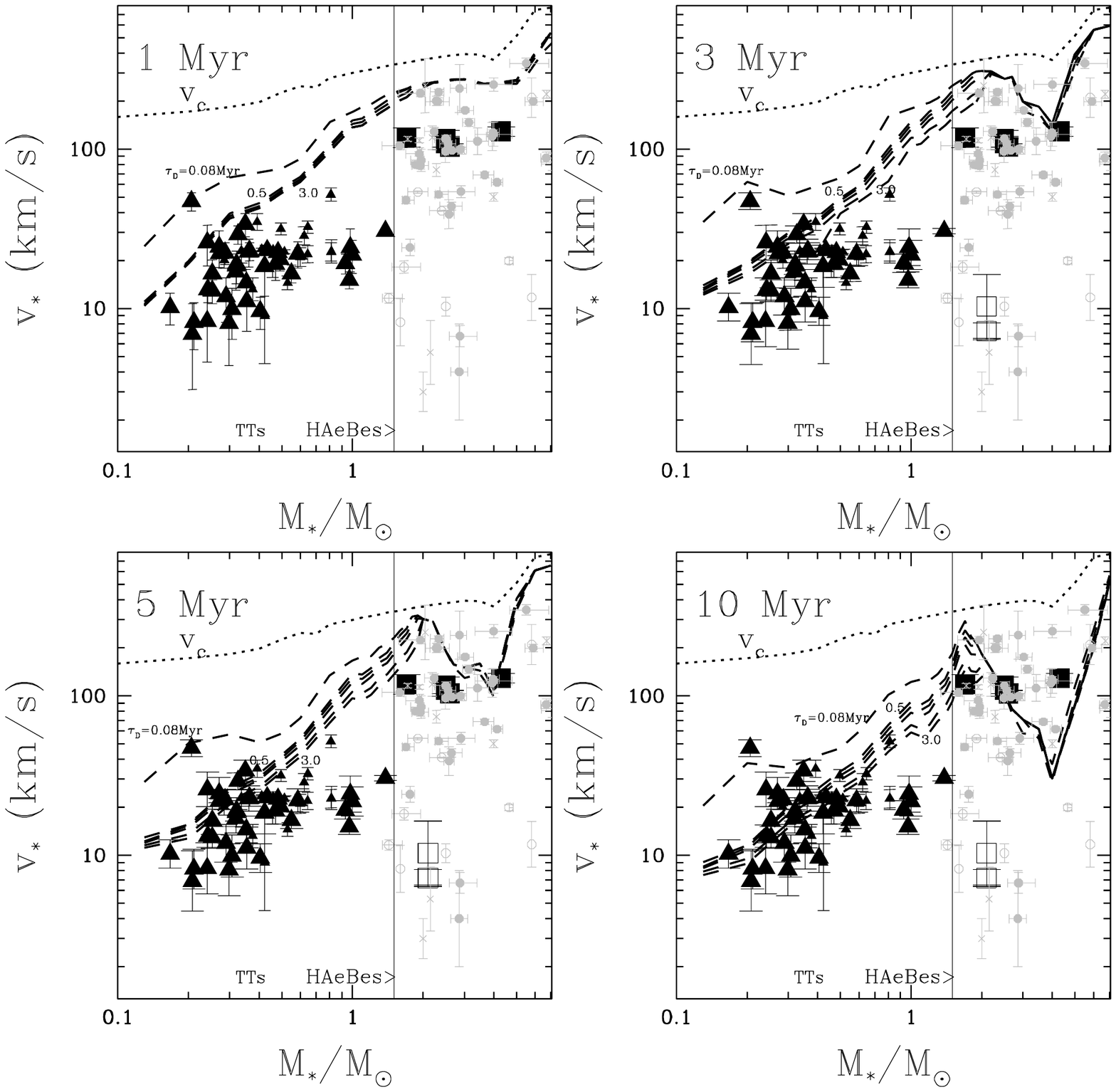}
\caption{Snapshot of the stellar rotation against mass at  $1$, $3$, $5$ and $10$Myr for $B_*=0.5$kG. The equatorial rotational velocities ($v_*$) are indicated with dashed lines for the six gas-disc lifetimes $\tau_{D}=0.08$, $0.5$, $0.7$,$1.0$, $2.0$ and  $3.0$Myr.   The break-up limit $v_{c}$  at the birth-line  is plotted with a dotted curve. Our $v$sin$i$ measurements  are indicated with  filled triangles in black (TTs) and filled squares in black (HAeBes).  Accretors among TTs are indicated with bigger triangles.   The complementary samples of \cite{cody2010} and \cite{alecian2013} are shown in gray. Binaries are indicated with empty circles while limit values of $v$sin$i$ with diagonal crosses. The vertical dashed  line marks the arrival to the MS while the horizontal dotted line  corresponds to median of rotational velocities of our HAeBes sample.  \label{fig:f12a}}
\end{figure*}

\begin{figure*}
\plotone{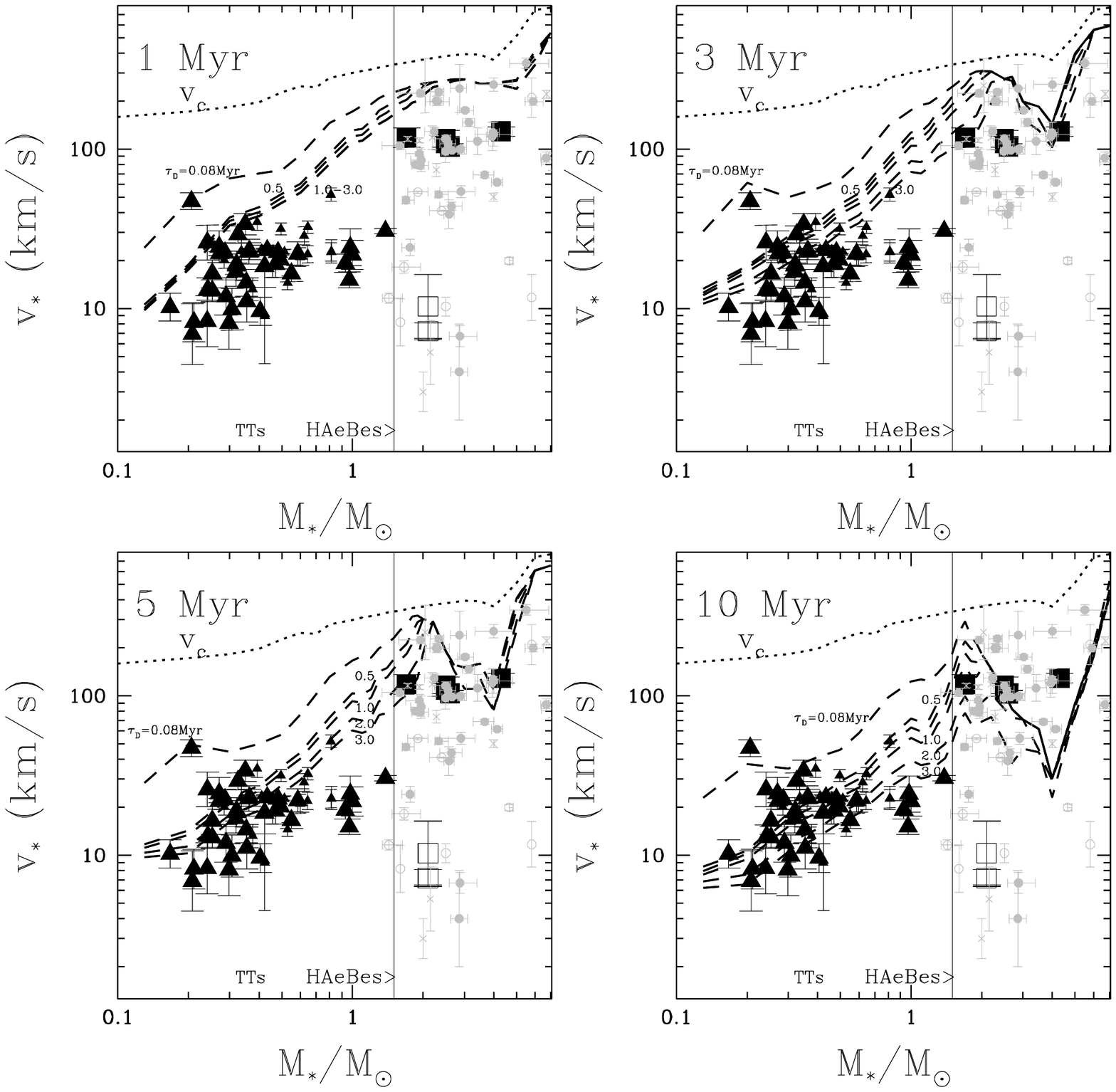}
\caption{Results for $B_*=1$kG in the same format as Figure  \ref{fig:f12a}. The spin-down is slightly larger and changes in velocity for distint $\tau_D$'s are noted, especially in the TTs zone.\label{fig:f12b}}
\end{figure*}

\begin{figure*}
\plotone{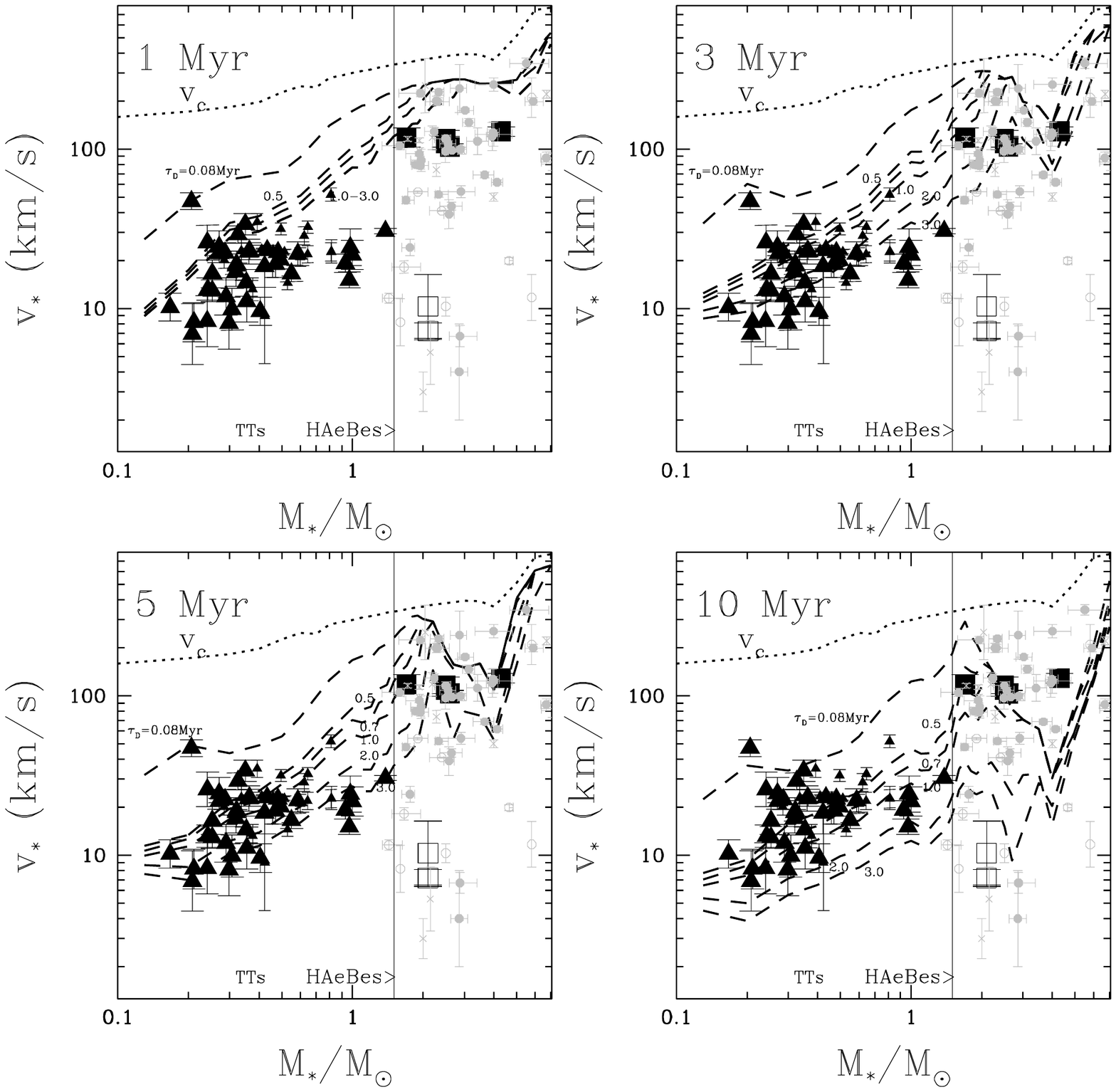}
\caption{Results for $B_*=2$kG in the same format as Figure  \ref{fig:f12a}. Spin down have progessively increased reaching values of 10\% at $10$ Myr. Despite of the best match with the data is obtained for this age, HAeBes remain unmattched. 
\label{fig:f12c}}
\end{figure*}

\begin{figure*}
\plotone{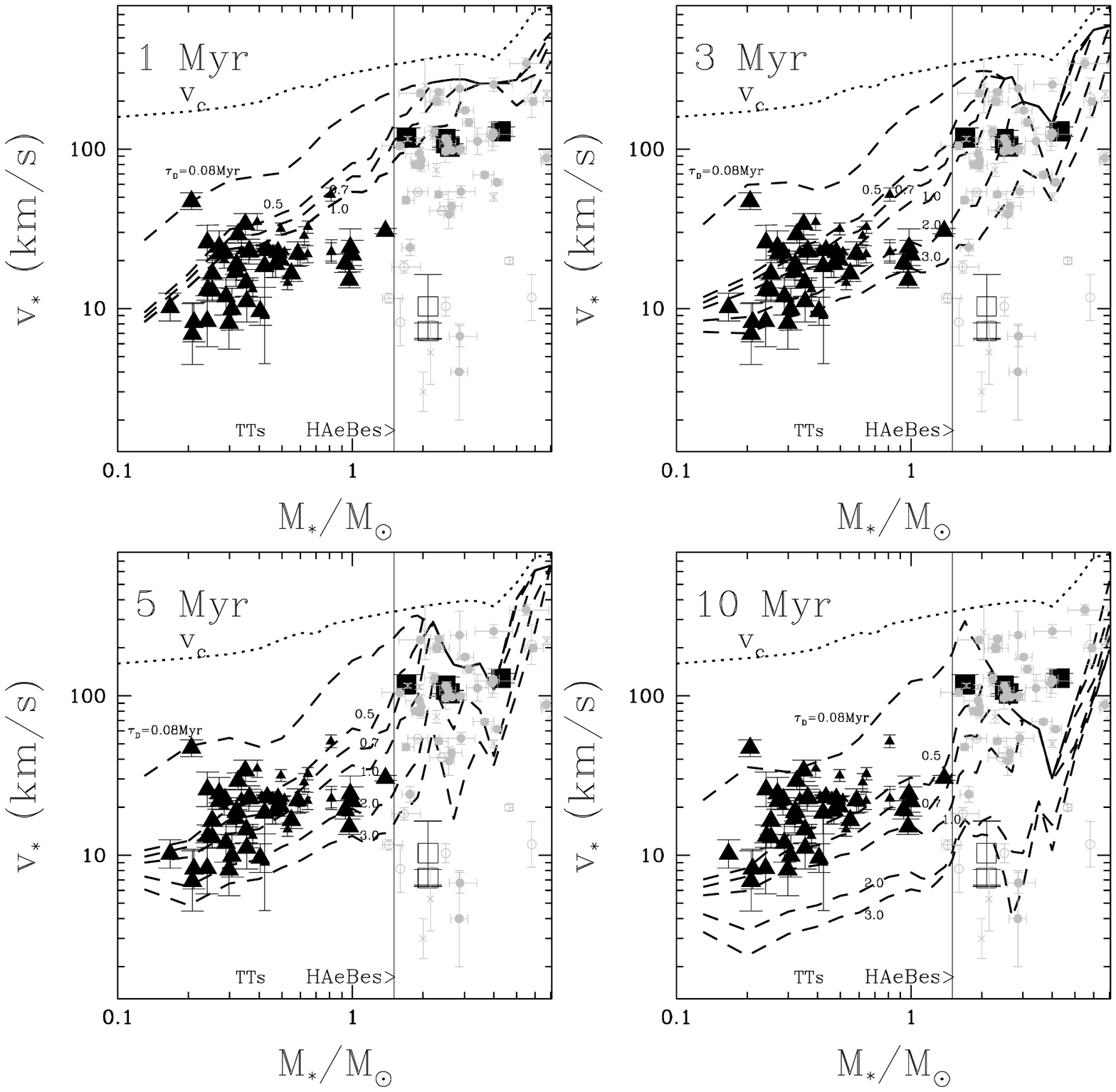}
\caption{Results for $B_*=3$kG in the same format as Figure  \ref{fig:f12a}. 
\label{fig:f12d}}
\end{figure*}

\begin{center}
\begin{figure*}
\begin{center}
\includegraphics[width=16.0cm, height=20.0cm]{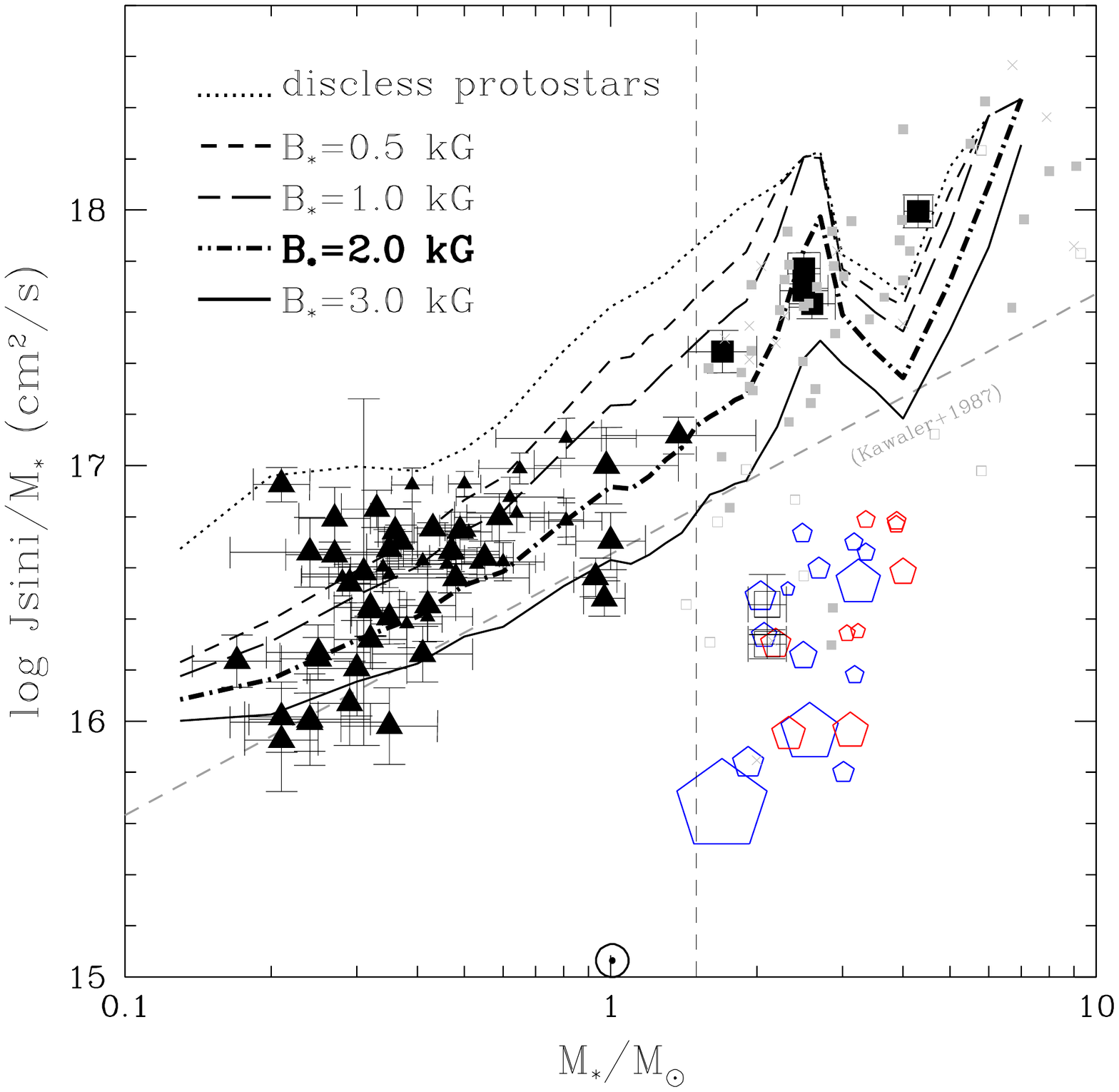}            
\caption{ Angular momentum as a function of mass for TTs in the $\sigma$-Orionis cluster (triangles) and  HAeBes in OB1 (rectangles in black).  The binary system HIP25258 is indicated with empty rectangles.    TTs with active accretion are plotted with large triangles in black while non-accretors with small-triangles. As complementary data for HAeBes we included with rectangles in gray, the sample studied by \cite{alecian2013} with open symbols representing  binaries and crosses limit values. The dashed grey line  represents the expected behaviour for normal stars on the main-sequence.  Pentagons correspond to the sample of Ap/Bp stars of  \cite{auriere2007} with size symbols proportional to the dipolar field strengths reported by the authors. Symbols in blue correspond to Ap stars younger than $10$ Myr whereas whereas older stars, ranging from $10$ up to $\sim300$ Myr, are indicated in red. Tracks (in black with different line-style), represent a snapshot of angular momentum at $3$ Myr for several magnetic field configurations  assuming that magnetic interaction between protostar  and its surrounding accretion disc started since the birth-line (see in the text).  Vertical dashed line represent the boundary between TTs and HAeBes. Angular momentum of the Sun is assumed to be $10^{15}$ $cm^2$ $s^{-1}$ for comparison purposes.}\label{fig:f13}
\end{center}
\end{figure*}
\end{center}

\end{document}